\documentclass[a4paper,11pt]{article}
\usepackage{jcappub}
\usepackage{lineno}
\pdfoutput=1

\usepackage{upgreek}
\usepackage{subcaption}
\usepackage{amsmath}
\usepackage{amssymb}

\title{The effect of quantum decoherence on inflationary gravitational waves}

\author[a,b]{Jessie de Kruijf}
\author[a,b,c]{Nicola Bartolo}
\affiliation[a]{Dipartimento di Fisica Galileo Galilei, Università di Padova, I-35131 Padova, Italy}
\affiliation[b]{INFN Sezione di Padova, I-35131 Padova, Italy}
\affiliation[c]{INAF-Osservatorio Astronomico di Padova, Italy}

\emailAdd{jessiearnoldus.dekruijf@phd.unipd.it, nicola.bartolo@pd.infn.it}

\abstract{
The theory of inflation provides a mechanism to explain the structures we observe today in the Universe, starting from quantum-mechanically generated fluctuations. However, this leaves the question of: how did the quantum-to-classical transition, occur? During inflation, tensor perturbations interact (at least gravitationally) with other fields, meaning that we need to view these perturbations as an open system that interacts with an environment. In this paper, the evolution of the system is described using a Lindblad equation, which describes the quantum decoherence of the system. This is a possible mechanism for explaining the quantum-to-classical transition. We show that this quantum decoherence during a de Sitter phase leads to a scale-dependent increase of the gravitational wave power spectrum, depending on the strength and time dependence of the interaction between the system and the environment. By using current upper bounds on the gravitational wave power spectrum from inflation, obtained from CMB and the LIGO-Virgo-KAGRA constraints, we find an upper bound on the interaction strength. Furthermore, we compute the decoherence criterion, which indicates the minimal interaction strength needed for a specific scale to have successfully decohered by the end of inflation. Assuming that the CMB modes have completely decohered, we indicate a lower bound on the interaction strength. In addition, this decoherence criterion allows us to look at which scales might not have fully decohered and could still show some relic quantum signatures. Lastly, we use sensitivity forecasts to study how future gravitational-wave detectors, such as LISA and ET, could constrain the decoherence parameter space. Due to the scale-dependence of the power spectrum, LISA could only have a very small impact. However, ET will be able to significantly improve our current constraints for specific decoherence scenarios. 

}

\begin{document}
\maketitle
\flushbottom

\section{Introduction} \label{sec: Introduction}

A period of accelerated expansion during the early universe, namely inflation \cite{STAROBINSKY198099,Guth:1980zm}, provides a mechanism to explain the origin of all structures in our Universe, namely the large-scale structure (LSS), and the Cosmic Microwave Background (CMB) anisotropies. Vacuum quantum fluctuations (of one or more scalar fields) were stretched on cosmological scales by accelerated expansion and (gravitationally) amplified during the accelerated expansion epoch and became primordial density fluctuations that eventually led to the formation of the LSS, which we observe today \cite{Mukhanov:1982nu,Starobinsky:1982ee,Bardeen:1983qw,Guth:1982ec}. Through the same mechanism, inflation also predicts the generation of a relic background of primordial gravitational waves (GWs)\cite{Starobinsky:1979ty,Rubakov:1982df,Allen:1987bk,1975JETP...40..409G, Fabbri:1983us, Abbott:1984fp}. This scenario and its predictions are in complete agreement with a variety of cosmological observations (see, e.g. \cite{Planck:2019kim, Planck:2018jri, 2020A&A...641A...6P}). 

However, the quantum origin of the perturbations also lead to new questions (see, e.g.,\cite{Perez:2005gh,Ashtekar:2020,Green_2020, Polarski:1995jg, Guth:1985ya}). We observe a classical universe so an important question is how the quantum-to-classical transition occurred in a cosmological context. At the moment, there is no definitive answer, although such an issue has been studied extensively \cite{ Kiefer:2008ku,Lesgourgues:1996jc, Kiefer:1998qe,Campo:2005sv,Burgess_2008, Sudarsky:2009za,Pinto-Neto:2013npa,Martin:2012pea,Das:2013qwa,Maldacena:2015bha, Martin:2015qta, Choudhury_2017, Choudhury_2022, Choudhury_2017_prim_cos, arani2024constrainingtensortoscalarratiobased}. One of the most important possible mechanisms to explain this transition is quantum decoherence, since it is a well-established and physically tested concept \cite{Brune:1996zz,Schlosshauer:2019ewh,Zurek:1981xq,Burgess_2008,Schlosshauer:2014pgr,Barvinsky:1998cq, Chandran_2024}. 

The most important aspect of quantum decoherence is that we are working with an open quantum system that interacts with its environment, not a closed one \cite{Zurek:1981xq, Joos1985TheEO, Schlosshauer_2005,10.1093/acprof:oso/9780199213900.001.0001, giulini2013decoherence, Rivas_2012}. 
There are two different types of environmental interactions that have been studied for cosmic decoherence. First, decoherence can occur between large- and small-scale modes of the same fluctuation fields. In this case, the short-wavelength modes act as the environment for the long-wavelength system \cite{Lombardo_2005, Burgess_2008, Nelson_2016, Gong_2019, Burgess_2023}. The second type of decoherence, and the one we will focus on in this work, is the decoherence arising from interactions between the inflationary field and an external field that represents the environment \cite{1801.09949, Martineau_2007, 2211.07598, Boyanovsky_2015, Liu_2016, burgess2024cosmicpuritylostperturbative}. The study of these field-field interactions is motivated by the likely presence of multiple fields in the early universe. 

The interaction between the system and the environment suppresses the off-diagonal elements of the reduced density matrix of the system. However, due to the time duration of the decoherence, the diagonal elements of the density matrix are also affected, which alters the probabilities associated with the final outcomes of measurements. In several works \cite{Martineau_2007, Burgess_2008, 1801.09949, 2211.07598, Boyanovsky_2015, 2018JCAP...06..037M} it is shown that, in fact, the statistical properties of the primordial perturbations are affected. These changes are shown to be essential if one accounts for the quantum-to-classical transition in the early universe, and observing them would give concrete proof of both inflation and decoherence during inflation. In addition, they carry observational signatures of quantum origin for primordial fluctuations.

Despite the fact that cosmic decoherence has been studied extensively, including in \cite{1801.09949, 2211.07598, Burgess_2008, Barvinsky:1998cq, Lombardo_2005, Nelson_2016, Liu_2016, Martineau_2007}, most of the literature has mainly focused on scalar perturbations. We extend on this by using the techniques presented in \cite{1801.09949, 2211.07598} and applying them to tensor perturbations from inflation (namely primordial GWs from inflation). In particular, we study the GW power spectra including a term induced by decoherence due to, e.g., an external field acting as an environment, and how this change in the power spectrum varies depending on several parameters. 

A further aim of this paper is to investigate how to constrain the decoherence process for gravitational waves by exploiting not only CMB information, but also the one coming from interferometers. This leads us also to the question of whether, and on which frequencies, the primordial inflationary gravitational waves might have preserved some relic signatures of their quantum origin.
Using the upper bound on the GW power spectrum, from CMB \cite{ 2018PhRvL.121v1301B, 2020A&A...641A...6P, 2021A&A...647A.128T, 2021PhRvL.127o1301A, 2022PhRvD.105h3524T, Galloni_2023, 2014A&A...571A..16P, Planck:2015fie,Planck:2018jri} and ground-based laser interferometers  \cite{PhysRevLett.118.121101, PhysRevD.104.022005, Abbott_2021}, one can find an upper bound on the  interaction strength between the system and the environment. This can be combined with the decoherence criterion, which indicates how mixed the state of the system has become due to the interaction with the environment (see, e.g.,~\cite{Burgess_2008}). This depends on the strength of the interaction and therefore assuming that CMB scales have fully decohered by the end of inflation gives us a lower bound on the interaction strength on CMB scales. However, very small scales (such as those probed by interferometers) might not have fully decohered at the end of inflation, meaning there would be some genuine quantum signatures left on those scales \cite{2021PhRvD.103d4017K, micheli2023quantum, 2023arXiv230202584N, Campo:2005sv,Martin2023, PhysRevD.61.024024, Banerjee_2023, Sharifian_2024, Parikh_2020, Parikh_2021, Parikh_2021_signatures, Lamine_2006, Choudhury_2017}. If these quantum signatures are observed, this would conclusively prove the quantum origin of primordial fluctuations, which is therefore of fundamental importance to the scientific community. In addition, using such an observation in combination with the decoherence criterion, would allow us to tighten the upper bounds on the interaction strength of the decoherence process.

Several future GW detectors are on the horizon, including at least two next-level ground-based laser interferometers, namely the Einstein Telescope \cite{Maggiore_2020, Branchesi_2023} and the Cosmic Explorer \cite{reitze2019cosmic, evans2021horizon}, and the space-based laser interferometer LISA \cite{LISA, Auclair_2023, Bartolo_2016}. For both ET and LISA we study which decoherence scenarios they could observe or, in case of no detection, exclude. Recently, the Pulsar Timing Array (PTA) method has been used to observe the nanohertz GW background \cite{Verbiest_2021, 2023ApJ...951L...8A, 2023arXiv230616226A, 2023ApJ...951L...6R, InternationalPulsarTimingArray:2023mzf, Xu_2023}. More research is needed to definitively state the origin of this signal, but it is nonetheless also a very promising step toward a better understanding of early universe physics.

This work is organized as follows. We start in Section \ref{sec: Lindblad equation} with a brief overview of the method applied in this work, mainly summarizing the approach introduced in \cite{1801.09949} (and used in ~\cite{2211.07598} for a first application to inflationary gravitational waves). In Section \ref{sec: Power spectrum} we derive the full expression for the GW power spectrum, including the additional decoherence term, and study how this changes when using different parameter values. Section \ref{sec: Constraints interaction strength} focuses on constraining the interaction strength between the system and the environment, both using observational data and the decoherence criterion. Moreover, we look at which decoherence scenarios could be probed with future detectors and we address the crucial question of at which scales (frequencies) the gravitational waves might not have fully decohered, thus leaving some relic quantum signatures on those scales. Finally, in Section \ref{sec: Conclusion} we summarize our results and discuss possible future work. In Appendix A we give some details on the computations performed to get the main results. In Appendix B we discuss and analyze the power-spectrum and decoherence arising from anther interaction term between tensor and scalar perturbations, already considered in~\cite{2211.07598}. Appendix C presents the equations governing the cross-spectra between different polarization states of gravitational waves, where we point out also the emergence of some interesting effects regarding the gravitational waves' polarization.



\section{Master equation}\label{sec: Lindblad equation}

\subsection{Lindblad equation}

During inflation, the primordial perturbations interact, at least gravitationally, with other degrees of freedom, possibly including the standard-model fields. Therefore, we consider inflationary tensor perturbations as an open system ``S", interacting with a given environment ``E", defined only by certain characteristics (within an effective approach). As we are only interested in the effect the environment has on the system, the environmental degrees of freedom are traced out during the derivation of the master equation for the system. This leads to the Lindblad equation that describes the evolution of the system. The full derivation of the Lindblad equation is described, e.g., in Appendix A of \cite{1801.09949}, but here we briefly recap the most important aspects, parameters, and assumptions. The Lindblad equation is derived using Born and Markov approximations \cite{Schlosshauer:2014pgr, Breuer:2003avm, Lindblad:1975ef, Gorini:1975nb}. The former means that we can take our equations to be perturbative in terms of the coupling. The latter means that the typical correlation time of the environment is much shorter than the time over which the system evolves (so the effect the system has on the environment is negligible). 

The total Hamiltonian, for both the system and the environment, in the Hilbert space $\mathcal{E} = \mathcal{E}_{S} \otimes \mathcal{E}_{E}$ is given by
\begin{equation}
    \hat{H} = \hat{H}_{\mathrm{S}} \otimes \hat{\mathcal{I}}_E + \hat{\mathcal{I}}_\mathrm{S} \otimes \hat{H}_\mathrm{E} +g\hat{H}_{\mathrm{int}},
\end{equation}
where $\hat{H}_{S} (\hat{H}_E) $ is the free Hamiltonian of the system (environment), $\hat{\mathcal{I}}_{\rm S} (\hat{\mathcal{I}}_{\rm E})$ the identity operator acting $\mathcal{E}_{S} (\mathcal{E}_{E})$, and $\hat{H}_{\mathrm{int}}$ is the interaction Hamiltoniain with a dimensionless coupling parameter $g$ characterizing the strength of interactions between the system and the environment. Assuming the interactions are local, we can write the interaction between the system and environment in the form\footnote{Notice that the interaction term (\ref{eq:interaction Hamiltonian}) is not of the form normally required to derive a Lindblad equation. However, in Appendix A \cite{1801.09949}, it is shown that the usual treatment can be generalized to an interaction term of the form (\ref{eq:interaction Hamiltonian}).}
\begin{equation}\label{eq:interaction Hamiltonian}
    \hat{H}_{int} (\eta) = \int \mathrm{d}^3 \boldsymbol{x} \hat{A}(\eta, \boldsymbol{x}) \otimes \hat{R}(\eta,\boldsymbol{x}),
\end{equation}
where $\Hat{A} (\Hat{R})$ is the local functional of the fields describing the system (environment) sector, and $\eta$ is the conformal time. 

The complete system is described by a density matrix $\hat{\rho}$, which at the initial time can be written as $\hat{\rho} (t_{in}) = \hat{\rho}_S (t_{in}) \otimes \hat{\rho}_E (t_{in})$, assuming that the system and the environment become entangled only at later times. As we are only interested in the evolution of the system, and we treat the environment in an effective way as unobservable degrees of freedom, we focus on the reduced density matrix
\begin{equation}
    \hat{\rho}_S (t)= \mathrm{Tr_E}[\hat{\rho}(t)],
\end{equation}
where the environment degrees of freedom have been traced out. This $\hat{\rho}_S (t)$ evolves according to the Lindblad equation\footnote{Also called Gorini-Kossakowski-Sudarshan-Lindblad equation.} \cite{Burgess_2008, 1801.09949, Gorini:1975nb, Shandera:2017qkg, Pearle_2012, Brasil_2013}
\begin{equation}\label{eq:Lindblad equation for rho sys} 
    \frac{\mathrm{d}\hat{\rho}_{\mathrm{S}}}{\mathrm{d}t} = i[\hat{\rho}_{\mathrm{S}},\hat{H}_{\mathrm{S}}] - \frac{\gamma}{2} \int \mathrm{d}^3\boldsymbol{x}\mathrm{d}^3\boldsymbol{y}  C_{R}(\boldsymbol{x},\boldsymbol{y}) [\hat{A}(\boldsymbol{x}),[\hat{A}(\boldsymbol{y}),\hat{\rho}_{\mathrm{S}}]],
\end{equation}
with $\gamma = 2g^2\eta_c$, where $\eta_c$ is the autocorrelation time of the environment. As $\gamma$ is in general time dependent, we adopt a power-law dependence on the scale factor, following \cite{1801.09949},
\begin{equation}\label{eq: gamma dependence on a}
    \gamma = \gamma_{\ast} \Bigl( \frac{a}{a_{\ast}} \Bigr)^p,
\end{equation}
where $p$ represents a free parameter, and $\ast$ refers to a reference time which is taken to be the time when the pivot scale $k_{\ast} = 0.05 \mathrm{Mpc}^{-1}$ crosses the Hubble radius (i.e. $k_{\ast} = a_{\ast}H_{\ast}$). 

Furthermore, $C_R(\boldsymbol{x}, \boldsymbol{y})$ is the same-time correlation function of the environment $\hat{R}$, defined by
\begin{equation}
    C_R (\boldsymbol{x},\boldsymbol{y}) = \mathrm{Tr_E}\big(\hat{\rho}_{E} \hat{R}(\boldsymbol{x})\hat{R}(\boldsymbol{y}) \big).
\end{equation}
Assuming that the environment is statistically homogeneous and isotropic, and that a single physical length scale $l_{E}$ is involved,  we take this to be a top-hat function, using \cite{1801.09949}
\begin{equation}\label{eq: Cr in phase space}
    C_R (\boldsymbol{x},\boldsymbol{y}) = \Bar{C}_{R}  \Uptheta \biggl( \frac{a|\boldsymbol{x}-\boldsymbol{y}|}{l_{\mathrm{E}}} \biggr),
\end{equation}
where $\Uptheta(x)$ is $1$ if $x<1$ and $0$ otherwise, and $\Bar{C}_R$ is a constant. In Fourier space this can be written as
\begin{equation}\label{eq: C tilde with sin cos}
    \Tilde{C}_R (k) = \sqrt{\frac{2}{\pi}} \frac{\Bar{C}_R}{k^3} \biggl[ \sin{\biggl( \frac{kl_{\mathrm{E}}}{a}\biggr)} - \frac{kl_{\mathrm{E}}}{a}\cos{\biggl( \frac{kl_{\mathrm{E}}}{a}\biggr)} \biggr],
\end{equation}
which can again be approximated by a top-hat function
\begin{equation}\label{eq:C with theta}
    \Tilde{C}_R (k) \simeq \sqrt{\frac{2}{\pi}} \frac{\Bar{C}_Rl_{\mathrm{E}}^3}{3a^3} \Uptheta \biggl( \frac{kl_{\mathrm{E}}}{a}\biggr),
\end{equation}
where the amplitude at the origin has been matched. 

Usually the Lindblad equation (\ref{eq:Lindblad equation for rho sys}) cannot be fully solved, therefore it is convenient to write it in terms of quantum expectation values 
\begin{equation}\label{eq:quantum exp values}
    \Big \langle \hat{O} \Big \rangle = \mathrm{Tr} (\hat{\rho}_{S}\hat{O})\, ,
\end{equation}
where $\hat{O}$ is an arbitrary operator acting in the Hilbert space of the system. This leads to a Lindblad equation of the form:
\begin{equation}\label{eq:Lindblad equation}
    \frac{\mathrm{d}\langle \hat{O}\rangle}{\mathrm{d}\eta} = \biggl \langle \frac{\partial\hat{O}}{\partial\eta} \biggr \rangle -i[\hat{O},\hat{H_S}] - \frac{\gamma}{2} \int \mathrm{d}^3\boldsymbol{x}\mathrm{d}^3\boldsymbol{y} C_R (\boldsymbol{x},\boldsymbol{y}) \bigl \langle [\hat{O},\hat{A}(\boldsymbol{x})],\hat{A}(\boldsymbol{y})] \bigr \rangle.
\end{equation}

\subsection{Tensor perturbations}\label{subsec: tensor perturbations}

In this work we study how inflationary tensor perturbations, $h_{ij}(\eta, \boldsymbol{x})$ are decohered by an environment. Focusing only on tensor perturbations, the perturbed (spatially flat) Friedmann-Lemaitre-Robertson-Walker (FLRW) metric \cite{Mukhanov1992TheoryOC} is given by
\begin{equation}
    \mathrm{d}s^2 =a^2(t)\big[ -\mathrm{d}\eta^2+(\delta_{ij}+h_{ij}(\eta, \boldsymbol{x})) \mathrm{d}x^i\mathrm{d}x^j \big],
\end{equation}
where $\eta$ is the conformal time, $\boldsymbol{x}$ the conformal spatial coordinate, and $a(t)$ the scale factor. Tensor perturbations are transverse ($\partial_ih_{j}^i = 0$), traceless ($h_i^i = 0$) and represent the dynamical degrees of freedom of the gravitational sector, meaning that they correspond to GWs. Analogously to \cite{2211.07598}, we use the Fourier expansion of tensor perturbations, 
\begin{equation}
    h_{ij}(\eta,\boldsymbol{x}) = \frac{1}{(2\pi)^{3/2}} \int \mathrm{d}^3\boldsymbol{k}h_{ij}(\boldsymbol{k})e^{i\boldsymbol{k}\cdot\boldsymbol{x}},
\end{equation}
with
\begin{equation}\label{eq: hij in terms of e}
    h_{ij}(\boldsymbol{k}) = \sum_{\lambda = +,\times} h_{\lambda}(\boldsymbol{k}) e^{\lambda}_{ij}(\boldsymbol{k}),     \quad e^{\lambda}_{ij}(\boldsymbol{k})e^{\lambda'}_{ij}(\boldsymbol{k}) = 2\delta^{\lambda\lambda'}.
\end{equation}

For canonical normalization, we use a tensor variable, similar to the Mukhanov-Sasaki variable \cite{Mukhanov1992TheoryOC,Lidsey:1995np} for scalar perturbations 
\begin{equation}\label{eq: v lambda}
    v_{\lambda} = \frac{aM_{pl}}{\sqrt{32\pi}}h_{\lambda}.
\end{equation}
At the perturbative linear order, the tensor perturbations will evolve according to the free Hamiltonian which, written in Fourier space, reads 
\begin{equation}\label{eq: definition H_v}
    \hat{H}_v = \frac{1}{2} \sum_{\lambda} \int \mathrm{d}^3 \boldsymbol{k} \Big[ \hat{p}_{\boldsymbol{k}}^{\lambda}\hat{p}_{-\boldsymbol{k}}^{\lambda} + \omega^2 \hat{v}_{\boldsymbol{k}}^{\lambda}\hat{v}_{-\boldsymbol{k}}^{\lambda} \Big] \quad \mathrm{with} \quad \omega^2 = k^2 - \frac{a''}{a},
\end{equation}
where $\hat{v}_{\boldsymbol{k}}$ is the Fourier transform of the tensor variable (\ref{eq: v lambda}), $\hat{p}_{\boldsymbol{k}}$ its conjugate momentum, i.e., $\hat{p}_{\boldsymbol{k}}^{\lambda} = \left( \hat{v}_{\boldsymbol{k}}^{\lambda} \right)^{\prime}$, and a prime denotes the derivative with respect to the conformatl time $\eta$.

For these tensor perturbations, we use the system interaction operator:
\begin{equation}
    \hat{A}(\eta,\boldsymbol{x}) = \partial_l h_{ij}(\eta,\boldsymbol{x}) \partial^l h^{ij}(\eta,\boldsymbol{x}).
\end{equation}
The focus is on the quadratic interaction because $h_{ij}$ is transverse, which means that at linear order the correction to the tensor power spectrum vanishes (see \cite{2211.07598}). Such an interaction term is inspired by the cubic tensor-scalar interactions arising in the Lagrangian of (single-field) inflation\footnote{ The cubic term of the action for two gravitons and a scalar is given by~\cite{Maldacena2003}: 
\begin{equation}
\label{GRcubic}
    S = \frac{\epsilon M_{pl}^2}{8} \int \mathrm{d}t \mathrm{d}^3x \bigg( a^3\zeta \Dot{h_{ij}}\Dot{h_{ij}} + a\zeta\partial_{l}h_{ij}\partial_{l}h_{ij} - 2a^3 \Dot{h_{ij}}\partial_{l}h_{ij}\partial_{l}\big(\nabla^2\big)^{-1} \Dot{\zeta} \bigg)\, ,
\end{equation}where the dot denotes a derivative with respect to time $t$. }, and it incorporates the spatial derivatives of the tensor fluctuation modes, naturally appearing in general relativity (GR). 

A specific model for the quantum decoherence of tensor perturbations following this approach has already been studied in \cite{2211.07598} (see also, e.g.  \cite{Brahma_2022, Ning_2023, Burgess_2023, Gong_2019}, which does not incorporate these spatial derivatives. However, from partial integration of the Lagrangian~(\ref{GRcubic}) it can be shown that both of these operators should be considered; therefore, we expand on this model a bit more in the Appendix \ref{sec:Daddi}. The exact interplay between the different decoherence models is left for future work. 

Inserting this system operator into (\ref{eq:Lindblad equation}), gives
\begin{equation}\label{eq:Lindblad first}
    \frac{\mathrm{d}\langle \hat{O}\rangle}{\mathrm{d}\eta} = \biggl \langle \frac{\partial\hat{O}}{\partial\eta} \biggr \rangle -i[\hat{O},\hat{H}_S] - \frac{\gamma\xi^2}{2} \int \mathrm{d}^3\boldsymbol{x}\mathrm{d}^3\boldsymbol{y} C_R (\boldsymbol{x},\boldsymbol{y}) \bigl \langle [\hat{O},\partial_l h_{ij}(\boldsymbol{x}) \partial^l h^{ij}(\boldsymbol{x})],\partial_a h_{bc}(\boldsymbol{y}) \partial^a h^{bc}(\boldsymbol{y})] \bigr \rangle,
\end{equation}
where $\xi^2$ is an expansion constant that serves to set the right dimensions. In Fourier space, and using Eq. (\ref{eq: v lambda}), this becomes\footnote{It may seem that there is a missing factor of $a^{-4}$ in the real part of (\ref{eq: Lindblad expressed in v}) but actually it is absorbed in the definition of $\gamma$ given in Eq.(\ref{eq: gamma dependence on a}), analogue to the approach in \cite{2211.07598}.}
\begin{equation}\label{eq: Lindblad expressed in v}
\begin{aligned}
    \frac{\mathrm{d}\langle \hat{O}\rangle}{\mathrm{d}\eta} = \biggl \langle \frac{\partial\hat{O}}{\partial\eta} \biggr \rangle -i[\hat{O},\hat{H}_v] - \frac{\beta^2\gamma}{2} \sum_{\lambda,\lambda'}\int &\mathrm{d}^3\boldsymbol{k}\mathrm{d}^3\boldsymbol{p_1}\mathrm{d}^3\boldsymbol{p_2} p_{1,l}(-k-p_1)^l p_{2,a} (k-p_2)^a \\
    &\cdot \Tilde{C}_R (|\boldsymbol{k}|) \bigl \langle [\hat{O},v_{p_1}^{\lambda}v_{-k-p_1}^{\lambda}],v_{p_2}^{\lambda'}v_{k-p_2}^{\lambda'}] \bigr \rangle,
\end{aligned}
\end{equation}
where we define 
\begin{equation}
    \beta = \frac{2\xi}{M_{pl}^4},
\end{equation}
being a coupling constant, used to maintain the analogy with \cite{1801.09949, 2211.07598}.



\section{Power spectrum of primordial gravitational waves}\label{sec: Power spectrum}

\subsection{Two-point correlation functions}

To obtain the GW power spectrum from Eq. (\ref{eq:Lindblad equation}), we insert two-point correlators of the form $\langle \hat{O} \rangle = \langle \hat{O}_{\boldsymbol{k}_1}\hat{O}_{\boldsymbol{k}_2} \rangle$ with $\hat{O}_{\boldsymbol{k}_i} = \hat{v}^{s}_{\boldsymbol{k}_i}$ or $\hat{p}^{s}_{\boldsymbol{k}_i}$, where $s$ refers to the polarization, finding: 
\begin{equation}\label{eq: first form two-point correlator functions}
\begin{aligned}
    \frac{\mathrm{d}\langle \hat{v}_{\boldsymbol{k}_1}^s \hat{v}_{\boldsymbol{k}_2}^s \rangle}{\mathrm{d}\eta} &= \langle \hat{v}_{\boldsymbol{k}_1}^s \hat{p}_{\boldsymbol{k}_2}^s \rangle + \langle \hat{p}_{\boldsymbol{k}_1}^s \hat{p}_{\boldsymbol{k}_2}^s \rangle \\
    \frac{\mathrm{d}\langle \hat{v}_{\boldsymbol{k}_1}^s \hat{p}_{\boldsymbol{k}_2}^s \rangle}{\mathrm{d}\eta} &= \langle \hat{p}_{\boldsymbol{k}_1}^s \hat{p}_{\boldsymbol{k}_2}^s \rangle -\omega^2 (k_2) \langle \hat{v}_{\boldsymbol{k}_1}^s \hat{v}_{\boldsymbol{k}_2}^s \rangle \\
    \frac{\mathrm{d}\langle \hat{p}_{\boldsymbol{k}_1}^s \hat{v}_{\boldsymbol{k}_2}^s \rangle}{\mathrm{d}\eta} &= \langle \hat{p}_{\boldsymbol{k}_1}^s \hat{p}_{\boldsymbol{k}_2}^s \rangle -\omega^2 (k_1) \langle \hat{v}_{\boldsymbol{k}_1}^s \hat{v}_{\boldsymbol{k}_2}^s \rangle \\
    \frac{\mathrm{d}\langle \hat{p}_{\boldsymbol{k}_1}^s \hat{p}_{\boldsymbol{k}_2}^s \rangle}{\mathrm{d}\eta} &=  -\omega^2 (k_2) \langle \hat{p}_{\boldsymbol{k}_1}^s \hat{v}_{\boldsymbol{k}_2}^s \rangle -\omega^2 (k_1) \langle \hat{v}_{\boldsymbol{k}_1}^s \hat{p}_{\boldsymbol{k}_2}^s \rangle  \\
    &+ \beta^2 \frac{4\gamma}{(2\pi)^{3/2}}\int \mathrm{d}^3 \boldsymbol{k} k_1^l (k+k_1)_lk_2^a(k_2-k)_a\Tilde{C}_R (|\boldsymbol{k}|) \langle \hat{v}^s_{\boldsymbol{k}+\boldsymbol{k}_1} \hat{v}^s_{\boldsymbol{k}_2-\boldsymbol{k}} \rangle\, .
\end{aligned}
\end{equation}

The system is solved through a perturbative expansion in $\gamma$, and  the environment correlator preserves statistical isotropy and homogeneity (see Eq. (\ref{eq: Cr in phase space})). This means we have a statistically homogeneous and isotropic solution of the form 
\begin{equation}\label{eq:general solution P OO}
    \bigl \langle \hat{O}_{\boldsymbol{k}_1} \hat{O}'_{\boldsymbol{k}_2} \bigr \rangle = (2\pi)^{3}P_{OO'}(k_1)\delta(\boldsymbol{k}_1+\boldsymbol{k}_2).
\end{equation}

In the last line of Eq. (\ref{eq: first form two-point correlator functions}), the last term shows that we sum over the indices $l$ and $a$. This allows us to take the dot product, giving
\begin{equation}\label{eq:dot product k}
    k_1^l(k+k_1)_l k_2^a(k_2-k)_a = (\boldsymbol{k}_1\cdot\boldsymbol{k})^2 +k_1^4 + 2k_1^2 (\boldsymbol{k}_1\cdot \boldsymbol{k}).
\end{equation}

Using (\ref{eq:dot product k}) and(\ref{eq:general solution P OO}), the system of equations (\ref{eq: first form two-point correlator functions}) can be written as\footnote{Since in Eq. (\ref{eq: first form two-point correlator functions}) all terms depend on the same polarization $s$, we drop it for simplification and do our computation for one polarization. When computing the final power spectrum we add a factor of $2$ to account for both polarizations. For more details on the polarization of the gravitational waves, see Appendix \ref{sec:polarization}.}
\begin{equation}\label{eq:all three diff of Pvv}
\begin{aligned}
    & \frac{\mathrm{d}P_{vv}(k)}{\mathrm{d}\eta} = P_{vp}(k) + P_{pv}(k) \\
    & \frac{\mathrm{d}P_{vp}(k)}{\mathrm{d}\eta} = \frac{\mathrm{d}P_{pv}(k)}{\mathrm{d}\eta} = P_{pp}(k) - \omega^2(k)P_{vv}(k) \\
    &\frac{\mathrm{d}P_{pp}(k)}{\mathrm{d}\eta} = -\omega^2(k) \Bigl( P_{pv}(k) + P_{vp}(k) \Bigr) \\
    &+ \beta^2 \frac{4\gamma}{(2\pi)^{3/2}}\int \mathrm{d}^3 \boldsymbol{k'} \Bigl( (\boldsymbol{k}\cdot \boldsymbol{k'})^2 + k^4 + 2k^2 (\boldsymbol{k}\cdot \boldsymbol{k'})\Bigr) \Tilde{C}_R (|\boldsymbol{k'}|) P_{vv}(|\boldsymbol{k}+\boldsymbol{k'}|).
\end{aligned}
\end{equation}

These equations can be combined into a single third-order equation for $P_{vv}$ 
\begin{equation}\label{eq:differntial equation P vv with S}
    P_{vv}''' + 4\omega^2(k)P_{vv}'+4\omega\omega'P_{vv} = S(\boldsymbol{k},\eta),
\end{equation}
where the source function $S(\boldsymbol{k}, \eta)$ is dependent on $P_{vv}$ and defined as
\begin{equation}\label{eq: source function}
    S(\boldsymbol{k},\eta) = \beta^2 \frac{8\gamma}{(2\pi)^{3/2}}\int \mathrm{d}^3 \boldsymbol{k'} \Bigl( (\boldsymbol{k}\cdot \boldsymbol{k'})^2 + k^4  + 2k^{2} (\boldsymbol{k'}\cdot \boldsymbol{k}) \Bigr) \Tilde{C}_R (|\boldsymbol{k'}|) P_{vv}(|\boldsymbol{k}+\boldsymbol{k'}|).
\end{equation} 
We obtain a source function similar to the one obtained by \cite{1801.09949} for a quadratic system operator, with a different coefficient in front and an additional dependence on $\boldsymbol{k}$, which arises from the derivatives in $\hat{A}$. Therefore, we continue using their approach, while accounting for this additional dependence. 
The source function depends on $P_{vv}$, making Eq. (\ref{eq:differntial equation P vv with S}) an integro-differential equation that couples all modes together. It is very difficult to solve this in full generality. However, since we are working at first order in $\gamma$, the free theory is used to calculate the source function $S(\boldsymbol{k}, \eta)$~\footnote{
Notice that this does not mean that the contributions coming from decoherence cannot be larger that the zero-order solution. The latter situation can happen e.g. if the system is coupled with an environment provided by an external, different degree of freedom. On the other hand, higher order contributions will be of $\mathcal{O}(\beta^4\sigma_{\gamma}^2)$ and higher. 
 Since $\beta^2\sigma_{\gamma}$ is small, these higher order contributions are expected to be smaller than the first order contribution. Moreover, as shown in Fig. \ref{fig:different sigma}, for lower values of the coupling, the decoherence induced term starts to dominate over the standard power spectrum at higher $k$. Therefore, we expect the effects of higher order contributions to arise at even higher $k$, possibly outside of the scale range considered in this work. For these reasons, we only focus on the first order contribution. Additionally, let us note that the form of the master equation we are using, i.e., the Lindblad equation, has been proven to provide a resummation at late-times (see \cite{2024arXiv240312240B}).}. The complete solution to (\ref{eq:differntial equation P vv with S}) is given by the sum of the homogeneous solution and a new term due to the interaction, namely\footnote{See \cite{1801.09949} for a more detailed explanation of this solution.}:
\begin{equation}\label{eq:solution diff equation Pvv}
    P_{vv}(k) =  |v_{\boldsymbol{k}}(\eta)|^2  + 2\int_{-\infty}^{\eta} S(\boldsymbol{k}, \eta')\mathfrak{Im}^2[v_{\boldsymbol{k}}(\eta') v_{\boldsymbol{k}}^{*}(\eta)] \mathrm{d}\eta'.
\end{equation}
Here $v_{\boldsymbol{k}}$ is Bunch-Davies normalized, meaning $P_{vv}$ matches the Bunch-Davies result in the infinite past if $\eta_{0} = -\infty$.  

\subsection{Computation of the source function}

The source function is computed explicitly, with the full derivation shown in Appendix \ref{sec:computation source function}, and here we provide just a short recap of the main steps and results. Starting by using a change of variables ($ \boldsymbol{k'} = \boldsymbol{p} - \boldsymbol{k} $) to perform the angular integration, we can write 
\begin{equation}\label{eq:begin integral}
\begin{aligned}
\int \mathrm{d}^3\boldsymbol{k'} \Bigl( (\boldsymbol{k}\cdot \boldsymbol{k'})^2 + k^4 + 2k^2(\boldsymbol{k}\cdot\boldsymbol{k'})\Bigr) \Tilde{C}_R (|\boldsymbol{k'}|) P_{vv}(|\boldsymbol{k}+\boldsymbol{k'}|) \\
=  \frac{\pi}{4k} \int_0^{\infty} \mathrm{d}p p  P_{vv}(p) \int_{(k-p)^2}^{(k+p)^2} \mathrm{d}z (k^2+p^2-z)^2 \Tilde{C}_R (\sqrt{z}).
\end{aligned}
\end{equation}
The integral over $z$ (which is defined as $z=k^2+p^2-2kp\cos{\theta}$, with $\theta$ the polar angle between {\bf k} and {\bf p}), is done using Eq. (\ref{eq: C tilde with sin cos}) and results in Eq. (\ref{eq:I first step}). 
To compute the second integral over $p$, we need to substitute the expression of the power spectrum, which can be divided into two cases. First, a UV part ($p>aH$, sub-Hubble scales) where $P_{vv}=(2p)^{-1}$, and an IR part ($p<aH$, super-Hubble scales) where $P_{vv}=(2p)^{-1}(-p\eta)^{-2}$. Note that we are neglecting slow-roll corrections, similarly to~\cite{1801.09949}, since the integral of the power spectrum over all modes appears in the source function. This means the slow-roll expansion around $k_{\ast}$ cannot be used consistently to describe the entire set of modes. A more complete calculation would have to be done on a model-by-model basis, which is beyond the scope of this work. Moreover, since the slow-roll parameters are very small and standard inflationary models predict the tensor spectral index to be slightly red-tilted, we expect them to have a negligible effect compared to the effect from quantum decoherence we show below. The UV part is regularized by adiabatic subtraction \cite{T_S_Bunch_1980, Markkanen_2018}, as is usually done. This is equivalent to neglecting the sub-Hubble part of the integral and limiting ourselves only to the super-Hubble perturbations by setting the upper bound of the integral over $p$ to be the comoving Hubble scale $-1/\eta$. 

Inserting the expression for $P_{vv}$ shows that the IR part is divergent and needs to be regularized by imposing an IR cut off $-1/\eta_{IR}$, which corresponds to the comoving Hubble scale at the onset of inflation. Using the known assumption\footnote{We assume the correlation length of the environment to be much smaller than the Hubble radius. This is true if $l_E \sim t_c$ since the derivation of the Lindblad equation requires $t_c \ll H^{-1}$, see Appendix A of \cite{1801.09949}. } $-l_E/(a\eta) \ll 1$, leads to 
\begin{equation}\label{eq:second integral}
    \int \mathrm{d}^3\boldsymbol{k'} \Bigl( (\boldsymbol{k}\cdot \boldsymbol{k'})^2 + k^4 + 2k^2(\boldsymbol{k}\cdot\boldsymbol{k'})\Bigr) \Tilde{C}_R (|\boldsymbol{k'}|) P_{vv}(|\boldsymbol{k}+\boldsymbol{k'}|) \simeq \frac{\pi k^2 \Tilde{C}_R(k)}{3\eta^2} \bigg( \frac{1}{\eta^2} - \frac{1}{\eta_{IR}^2} \bigg).
\end{equation}
Combining this expression (\ref{eq:second integral}) with (\ref{eq: source function}), the source function is given by
\begin{equation}\label{eq:source function computed}
    S(k,\eta) = \frac{2k^2\beta^2\gamma \Tilde{C}_R(k)}{3\eta^2} \sqrt{\frac{2}{\pi}} \bigg( \frac{1}{\eta^2} - \frac{1}{\eta_{IR}^2} \bigg).
\end{equation}
When comparing this result with the scalar perturbation case (\cite{1801.09949}), we again clearly see an additional $k^4$ dependence, as seen in Eq. (\ref{eq:dot product k}), this is due to the partial derivatives in $\hat{A}$.

\subsection{Power spectrum}

Finally, the power spectrum (\ref{eq:solution diff equation Pvv}) is obtained by inserting this source function (\ref{eq:source function computed}), together with the approximation (\ref{eq:C with theta}) and the parameterization (\ref{eq: gamma dependence on a}), giving
\begin{equation}
\begin{aligned}
    P_{vv} =  |v_{\boldsymbol{k}}(\eta)|^2  + \frac{8\gamma_{\ast}\beta^2l_E^3 \Bar{C}_R k^2\eta_{\ast}^{p-3}}{9\pi a_{\ast}^3}  \int_{-\infty}^{\eta} &\mathrm{d}\eta'  (\eta')^{1-p}\,  \Uptheta \bigg( \frac{kl_E}{a} \bigg) \bigg( \frac{1}{(\eta')^2} - \frac{1}{\eta_{IR}^2} \bigg) \\
    &\cdot \mathfrak{Im}^2[v_{\boldsymbol{k}}(\eta') v_{\boldsymbol{k}}^{\ast}(\eta)].
\end{aligned}
\end{equation}
To compute the integral, we start from the last term, using the explicit form of the Bunch-Davies normalized mode function
\begin{equation}\label{eq: v definition}
    v_{\boldsymbol{k}}(\eta) = \frac{1}{2}\sqrt{\frac{\pi}{k}} \sqrt{-k\eta} e^{-i\pi/2 (\nu+1/2)} H_{\nu}^{(2)}(-k\eta),
\end{equation}
with $H_{\nu}^{(2)}(z)$ the Hankel function of the second kind of order $\nu$, using $\nu \equiv 3/2$ as we neglect slow-roll corrections Inserting this\footnote{Decomposing the Hankel function into real and imaginary parts, $H_{\nu}^{(2)}(z) = J_{\nu}(z) - i Y_{\nu}(z)$, and making use of the relation $Y_{\nu}(z)=[J_{\nu}(z)\cos{(\nu\pi)} - J_{-\nu}(z) ]/[\sin{(\nu\pi)}] $.} and calculating the integral gives
\begin{equation}\label{eq:Pvv first expression}
\begin{aligned}
    P_{vv}(k) =  |v_{\boldsymbol{k}}(\eta)|^2 &+\frac{\pi\gamma_{\ast}\beta^2l_E^3 \Bar{C}_R }{18 a_{\ast}^3 \sin{(\nu\pi)}^2} k^3 (-k\eta_{\ast})^{p-3}(-k\eta)JI(k, \eta, \nu),
\end{aligned}
\end{equation}
where we use 
\begin{equation}
    JI(k, \eta, \nu) =  J_{-\nu}^2(-k\eta)I_1(\nu, k, \eta) -2J_{-\nu}(-k\eta)J_{\nu}(-k\eta)I_2(\nu, k, \eta) + J_{\nu}^2(-k\eta)I_1(\nu, k, \eta) ,
\end{equation}
and the integrals $I_1$ and $I_2$ are defined by
\begin{equation}\label{eq:I1,I2 definition}
\begin{aligned}
    &I_1(\nu, k, \eta) \equiv \int_{-k\eta}^{(H_{\ast}l_E)^{-1}} \mathrm{d}z z^{2-p} \bigg( \frac{1}{z^2} - \frac{1}{(-k\eta_{IR})^2} \bigg) J_{\nu}^2(z), \\
    &I_2(\nu, k, \eta) \equiv \int_{-k\eta}^{(H_{\ast}l_E)^{-1}} \mathrm{d}z z^{2-p} \bigg( \frac{1}{z^2} - \frac{1}{(-k\eta_{IR})^2} \bigg) J_{\nu}(z)J_{-\nu}(z).
\end{aligned}
\end{equation}
The upper bound in these integrals corresponds to the time when the wavelength $a/k$ of the co-moving mode under consideration $k$ crosses the correlation length of the environment $l_E$, at leading order, this is $-k\eta_E = (H_{E}l_E)^{-1} \simeq (H_{\ast}l_E)^{-1}$ (neglecting slow-roll corrections). 

The final step is to convert (\ref{eq:Pvv first expression}) to the total dimensionless GW power spectrum, using
\begin{equation}\label{eq:definition dimensionless PS}
    \mathcal{P}_T(k) = \frac{k^3}{2\pi^2} \frac{32\pi \cdot 2P_{vv}}{M_{pl}^2a^2} = \mathcal{P}_{T|{\mathrm{stan}}} [1+\Delta \mathcal{P}_T(k)],
\end{equation}
with
\begin{equation}\label{eq:PT standard from 1+delta}
    \mathcal{P}_{T|{\mathrm{stan}}} \simeq \frac{16}{\pi} \bigg( \frac{H}{M_{pl}} \bigg)^2,
\end{equation}
and 
\begin{equation}\label{eq:definition delta P}
    \Delta \mathcal{P}_T (k)= \frac{2\int_{-\infty}^{\eta} S(\boldsymbol{k}, \eta')\mathfrak{Im}^2[v_{\boldsymbol{k}}(\eta') v_{\boldsymbol{k}}^{*}(\eta)] \mathrm{d}\eta'}{ |v_{\boldsymbol{k}}(\eta)|^2}.
\end{equation}
In Eq. (\ref{eq:definition dimensionless PS}) we add a factor of two to account for both polarizations of the gravitational waves.
Inserting Eq. (\ref{eq:Pvv first expression}) and (\ref{eq: v definition}), we obtain
\begin{equation}\label{eq: PT full}
\begin{aligned}
    \mathcal{P}_{T}(k) = \mathcal{P}_{T|\mathrm{stan}} \Bigg( 1 + \frac{2 \beta^2 \sigma_{\gamma}}{9 \sin^2(\nu\pi)} \bigg( \frac{k}{k_{\ast}} \bigg)^{p+1} \frac{JI(k, \eta, \nu)}{[J_{\nu}^2(-k\eta) + Y_{\nu}^2(-k\eta)]} \Bigg).
\end{aligned}
\end{equation}

The $k^{p+1}$ again indicates a shift of $k^4$, w.r.t. the scalar perturbation case \cite{1801.09949}. For the purposes of this work, we mainly focus on a specific range of scales, namely scales that can be probed with ground-based interferometers up to scales probed by the CMB.
Furthermore, in these expressions, $N-N_{IR} \equiv \ln{(\eta_{IR}/\eta)}$ denotes the number of $e$-folds elapsed since the onset of inflation and $N-N_{\ast} \equiv \ln{(\eta_{\ast}/\eta)}$ denotes the number of $e$-folds elapsed since the pivot scale $k_{\ast}=0.05\mathrm{Mpc}^{-1}$ crosses out of the Hubble radius. We have also introduced the dimensionless coefficient
\begin{equation}
    \beta^2\sigma_{\gamma} \equiv  \frac{l_E^3\Bar{C}_R\gamma_{\ast}k_{\ast}^{4}\beta^{2}}{a_{\ast}^3},
\end{equation}
that characterizes the strength of the interaction with the environment. We chose to use this notation to clearly distinguish between the cases presented in \cite{1801.09949, 2211.07598}, but we will treat this as one parameter.   

\begin{figure}%
    \includegraphics[width=\textwidth]{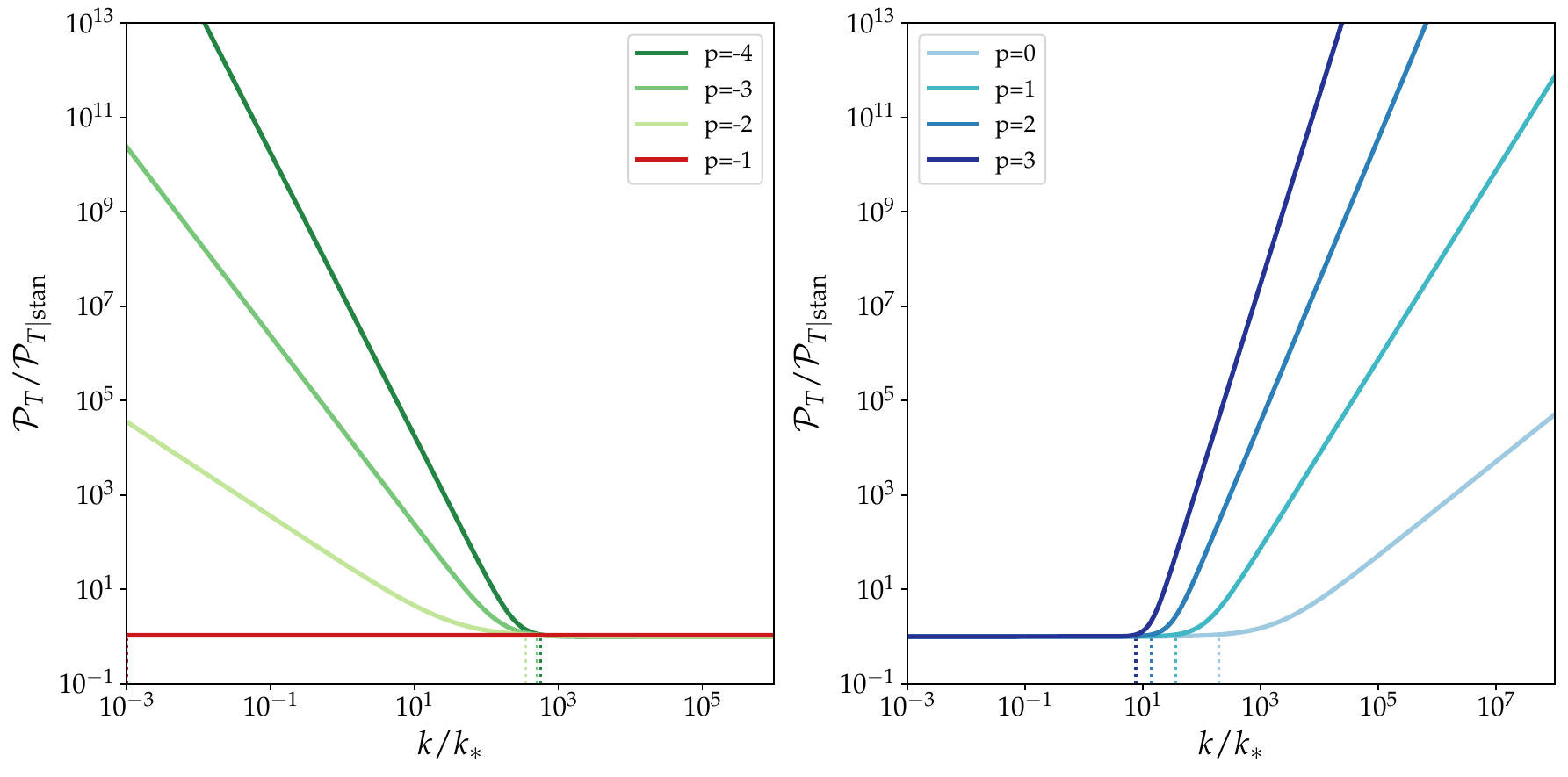}
    \caption{The GW power spectrum at the end of inflation, including the effects of decoherence. The colors represent different values of $p$. The dotted lines correspond to the transition scale $k_t$, defined by $\Delta P_{k_t}=1$. Assuming de Sitter inflation, the other model parameters are chosen to be $H_{\ast}l_E = 10^{-3}$, $\beta^2\sigma_{\gamma} = 10^{-3}$, $\Delta N_{\ast} = N_{end} - N_{\ast} = 50$ and $N_T = N_{end} - N_{IR} = 10^{4}$. }%
    \label{fig:Power spectrum}%
\end{figure}

The GW power spectrum is shown in Figure \ref{fig:Power spectrum}. The other model parameters are chosen to be $H_{\ast}l_E = 10^{-3}$, $\beta^2\sigma_{\gamma} = 10^{-3}$, $\Delta N_{\ast} = N_{end} - N_{\ast} = 50$ and $N_T = N_{end} - N_{IR} = 10^{4}$, similarly to \cite{1801.09949}, to simplify comparisons between the scalar and tensor case. Unless specifically stated otherwise, we use these values in all our figures. Let us emphasize that on small scales there is a cut-off point (at $-k\eta = (H_{\ast}l_E)^{-1}$) of the power spectrum. This is because a mode must have crossed the correlation length of the environment to be affected by the environment \cite{1801.09949}. This provides a natural cut-off mechanism depending on the value of the parameters. In the derivation of the Lindblad equation, we assume $H_{\ast}l_E \ll 1$, which means that the cut-off point is on scales outside of the range considered in this work (for the number of e-folds since the pivot scale entered the horizon chosen to be 50). Consequently, we will leave constraining the value of $H_{\ast}l_E$ for future work. 

Figure \ref{fig:Power spectrum} clearly shows that the tensor power spectrum can be divided into two distinct regions: one where the standard, (almost) scale-invariant tensor power spectrum dominates and the other characterized by a strong increase from decoherence. The only exception is the case $p=-1$, where we see that the contributions from decoherence effects remain scale-invariant. The dotted line indicates the scale $k_t$ where the transition between the two branches occurs, such that $\Delta P_{k_t}=1$. We can also clearly distinguish three different scenarios for the effect that decoherence has on the power spectrum,  depending on the value $p$ (see Eq. (\ref{eq: gamma dependence on a})). The first case, $p<-1$, only changes the power spectrum on large scales. Second, the case $p=-1$ results in a scale-invariant decoherence contribution to the power spectrum. And finally, for $p>-1$, decoherence affects only small scales. This shows that the main dependence of the power spectrum (\ref{eq: PT full}) on $p$ is encapsulated in the $k^{p+1}$ term, but note that the integrals inside the $JI(k, \eta,\nu)$ also depend on $p$.

\begin{figure}%
    \includegraphics[width=\textwidth]{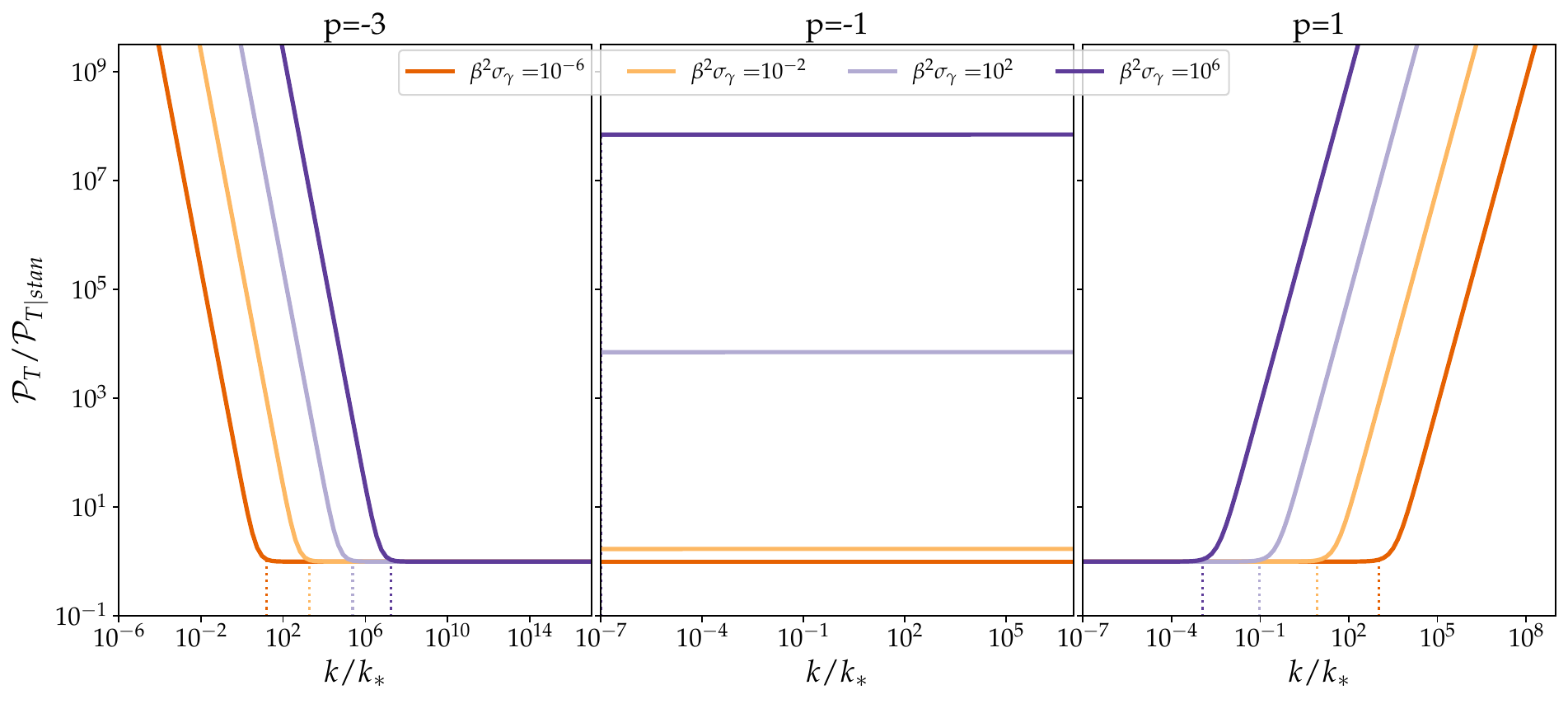}
    \caption{The GW power spectrum at the end of inflation, including the effects of decoherence. The colors represent different values of $\beta^2\sigma_{\gamma}$, shown for three different values of $p$, each corresponding to a different scenario ($p<-1, p=-1, p>-1$). The dotted lines correspond to the transition scale $k_t$, defined by $\Delta P_{k_t}=1$. Assuming de Sitter inflation, the other model parameters are chosen to be $H_{\ast}l_E = 10^{-3}$, $\Delta N_{\ast} = N_{end} - N_{\ast} = 50$ and $N_T = N_{end} - N_{IR} = 10^{4}$.}%
    \label{fig:different sigma}%
\end{figure}

One of the parameters that we will focus more on is the interaction strength, so in Fig. \ref{fig:different sigma} we illustrate the effect that changing this parameter has on the tensor power spectrum for different values of $p$. It shows that changing the value of the interaction strength changes the transition scale $k_t$, but does not affect the slope of our increase induced by decoherence. This behaviour is clear if one recalls that the main dependence of the decoherence-induced contribution to the tensor power spectrum is dictated by $(k/k_*)^{p+1}$ and that such contribution is proportional to the coupling strength $\beta^2 \sigma_\gamma$. Figure \ref{fig:different sigma} illustrates that the change in the power spectrum represents a potential signature of quantum decoherence and, therefore, also of the quantum nature of inflationary gravitational waves. This signature can be probed by present, and future, CMB, and (direct) interferometric observations, allowing us to put significant constraints on the interaction strength. 

\section{Constraints on the interaction strength}\label{sec: Constraints interaction strength}

From Eq. (\ref{eq: PT full}) it is clear that the full GW power spectrum depends on three quantities: $\mathcal{P}_{T|\mathrm{stan}}$, $\beta^2\sigma_{\gamma}$ and $p$. In order to constrain these parameters, we will make use of both observational constraints and of properties of the decoherence process. 

\subsection{Observational constraints}\label{subsec: Observational constraints}

There are several works on constraining the primordial GW power spectrum \cite{ 2018PhRvL.121v1301B, 2020A&A...641A...6P, 2021A&A...647A.128T, 2021PhRvL.127o1301A, 2022PhRvD.105h3524T, Galloni_2023, 2014A&A...571A..16P, Planck:2015fie,Planck:2018jri} , however, almost all of them are limited to observations at CMB scales. With direct detections of GWs by the LIGO, Virgo and KAGRA Collarboration (LVK) \cite{PhysRevLett.118.121101, PhysRevD.104.022005, Abbott_2021} new observational scales has become available. In this work we will use the results found in \cite{Galloni_2023} to constrain the primordial GW power spectrum, since it represents the most updated data analysis that leaves the spectral tensor index as a free parameter. In \cite{Galloni_2023} the tensor spectrum is parameterized as a power law, of the form
\begin{equation}\label{eq:power law observations}
    \mathcal{P}_{T|\mathrm{obs}}(k) = r_{\Tilde{k}}A_{s,\Tilde{k}} \bigg( \frac{k}{\Tilde{k}} \bigg)^{n_{T|\rm{obs}}},
\end{equation}
where the subscript $\Tilde{k}$ indicates that the parameters are evaluated at this pivot scale (corresponding to the pivot scale used in the data analysis), $r=A_{t}/A_{s}$ is the tensor-to-scalar perturbation ratio, $A_{s} (A_{t})$ is the amplitude of scalar (tensor) perturbations, and $n_{T|\rm{obs}}$ is the tensor spectral index considered in~\cite{Galloni_2023}. Assuming this simple power law, \cite{Galloni_2023} finds $r_{0.01} < 0.028$ and $-1.37 < n_{T|\rm{obs}} < 0.42$ at $95\%$ CL. As the cosmological GW background (CGWB) has not definitively been observed, this is an upper bound on the primordial tensor power spectrum. In addition to this, they also look for constraints using the CMB data and the 12.5 years NANOGrav collaboration results \cite{2020ApJ...905L..34A}, giving $r<0.033$ and $0.47 < n_{T|\rm{obs}} < 0.85$ at $95\%$ CL. This is inconsistent with the results obtained when using the LVK data and therefore will not be used in this work. Indeed, this inconsistency indicates that more complicated models for the power spectrum need to be considered when trying to account simultaneously of the Pulsar Timing Arrays and LVK data.  

\begin{figure}%
    \includegraphics[width=\textwidth]{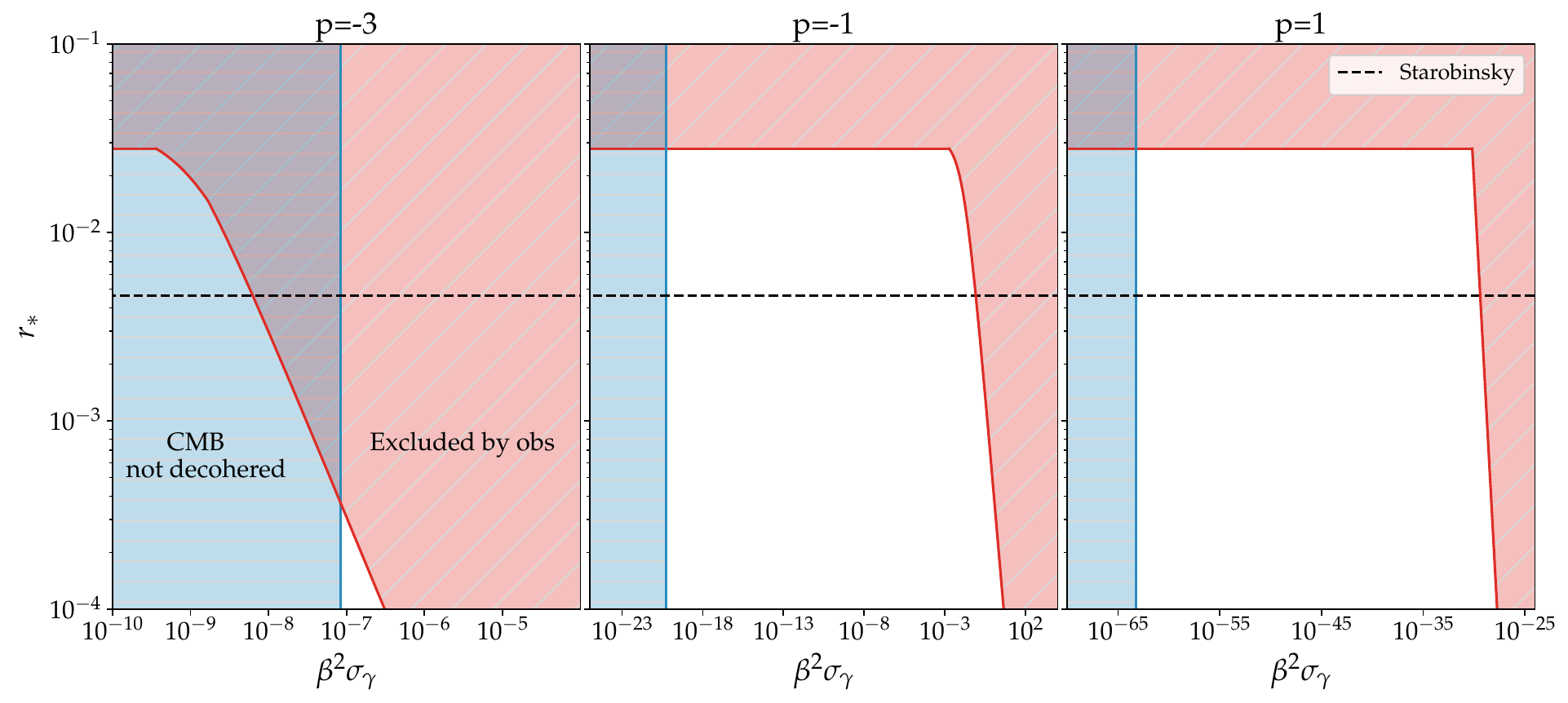}
    \caption{The red area shows the values of $\beta^2\sigma_{\gamma}$ excluded by observations done at scales ranging from CMB to LVK, shown for a range of $r_{\ast}$ indicating the amplitude of the spectrum at $k_{\ast}=0.05 \rm{Mpc^{-1}}$ and for three different values of $p$, each corresponding to a different scenario ($p<-1, p=-1, p>-1$). The blue region indicates the scenarios excluded due to assuming the CMB scales have completely decohered. For reference the Starobinsky model of inflation ($r_{\ast}=0.00461$) is shown. Assuming de Sitter inflation, the other model parameters are chosen to be $H_{\ast}l_E = 10^{-3}$, $\Delta N_{\ast} = N_{end} - N_{\ast} = 50$ and $N_T = N_{end} - N_{IR} = 10^{4}$. }%
    \label{fig:allowed region for sigma}%
\end{figure}

In order to constrain the interaction strength, we parameterize (\ref{eq:PT standard from 1+delta}) using 
\begin{equation}\label{eq:Pt standard}
    \mathcal{P}_{T|\mathrm{stan}} = r_{\ast}A_{s,k_{\ast}},
\end{equation}
where this is evaluated at $k_{\ast}=0.05 \mathrm{Mpc}^{-1}$, and we assume the standard power spectrum is scale-invariant ($n_{T|\mathrm{stan}}=0$), since the tensorial spectral index is expected to be very small in slow-roll inflation ($n_{T|\mathrm{stan}} = -2 \epsilon$ with $\epsilon \ll1$), and, as remarked above, we expect its effect to be negligible w.r.t. to the contributions coming from quantum decoherence. In order to consider more complex models, the $\mathcal{P}_{T|\mathrm{stan}}$ would simply have to be replaced with a more complex power spectrum. 
On cosmological scales ranging from LVK to CMB we require:
\begin{equation}\label{eq: obs constraint} 
    \mathcal{P}_{T}(k) < \mathcal{P}_{T|\mathrm{obs}}(k).
\end{equation}
Specifically, we assume the power law (Eq. (\ref{eq:power law observations})) holds only on these specific scales (LVK to CMB). Therefore, the upper bound on the power spectrum leads to an upper bound for the interaction strength, namely
\begin{equation}\label{eq:upper limit sigma from obs}
    \beta^2\sigma_{\gamma} < \Bigg[ \frac{r_{\Tilde{k}}}{r_{\ast}} \bigg( \frac{k}{\Tilde{k}} \bigg)^{n_{T|\rm{obs}}} -1 \Bigg] \cdot \Bigg[\frac{2}{9 \sin^2(\nu\pi)} \bigg( \frac{k}{k_{\ast}} \bigg)^{p+1} \frac{JI(k, \eta, \nu)}{[J_{\nu}^2(-k\eta) + Y_{\nu}^2(-k\eta)]} \Bigg]^{-1},
\end{equation}
where we use $A_{s,\Tilde{k}} \simeq A_{s,k_{\ast}} \simeq 2.1 \cdot 10^{-9} $ \cite{2020A&A...641A...6P, Galloni_2023} \footnote{This assumption automatically implies we require $r_{\ast}<r_{\Tilde{k}}$, in order to satisfy Eq. (\ref{eq: obs constraint}).}. Figure \ref{fig:allowed region for sigma} shows the highest values of $\beta^2\sigma_{\gamma}$ allowed by observations for each $r_{\ast}$ (red-shaded region), shown for different values of $p$. For reference, we show the Starobinsky model of inflation \cite{STAROBINSKY198099}, corresponding to $r_{\ast}=0.00461$ \cite{LiteBIRD:2022cnt}. This is the observational goal of future CMB surveys like LiteBIRD \cite{LiteBIRD:2022cnt}, which we expand on in Section \ref{subsec: Future obs}. Figure \ref{fig:allowed region for sigma} also shows for which scenarios the CMB scales have completely decohered, as discussed in Section \ref{subsec: Decoherence criterion}. It does not come as a surprise that there are two features that are clearly evident. On the one hand, the observational constraints (\ref{eq: obs constraint}) impose an upper bound on the interaction strength between the GW fluctuations modes (the system) and the environment, in such a way that the contributions to the power spectrum from the decoherence process are not too large to make the tensor power spectrum conflict with present observations. On the other hand, imposing that GWs have fully decohered on CMB scales imposes necessarily a lower bound on such coupling strength, because we are requiring the decoherence process to be sufficiently effective on those scales. Figure \ref{fig:allowed region for sigma} shows how different the allowed interaction strength is, depending on the scenario for $p$. This is because of the scales on which we have constraints. Figure \ref{fig:Power spectrum} shows that for a specific value of $\beta^2\sigma_{\gamma}$, all $p$ scenarios have their transition scale at (almost) the same value. Because of the way the interaction strength affects the power spectrum (see Figure \ref{fig:different sigma}), the constraints on scenarios with $p<-1$ (which mainly depend on the CMB observations) are much weaker (in terms of the magnitude of the coupling strength allowed) than those for scenarios with $p>-1$ (which mainly depend on the LVK upper bound). This is because in the latter case the tensor power spectrum increases with the scale, over many orders of magnitude (from CMB to LVK scales), thus shifting the transition scales to much smaller scales and in turn pushing the coupling strength to be very small. The same analysis has been applied to the model presented in \cite{2211.07598}, for which the results are shown in Appendix \ref{sec:Daddi}. We similarly find that the scenarios with an increase on small scales are much better constrained.

Recently, the Pulsar Timing Array method was used to find evidence for a GW background \cite{Verbiest_2021, 2023ApJ...951L...8A, 2023arXiv230616226A, 2023ApJ...951L...6R, InternationalPulsarTimingArray:2023mzf,Xu_2023}, although it has not yet
been definitively determined whether it is of cosmological or astrophysical origin (see, e.g., \cite{Nano_new_physics,Figueroa:2023zhu, bian2023gravitationalwavesourcespulsar, wang2022novelphysicsinternationalpulsar}. Due to the frequency and amplitude of the signal, it is not possible that this signal aises from the quantum decoherence scenarios analyzed here. As any decoherence scenario corresponding to this observed signal would have already been observed in the CMB or with LVK, because the cut-off of the power spectrum is on scales smaller than LVK (due to the requirement that $H_{\ast}l_{E}\ll 1$). 

In addition to the methods mentioned above, Big Bang Nucleosynthesis (BBN) can also be used to constrain the gravitational wave power spectrum, especially on small scales \cite{Maggiore_2000, Boyle_2008}.This bound is much weaker than the ones provided by LVK and CMB, but because it extends to much higher frequencies, it could still be useful to constrain quantum decoherence (depending on the environmental correlation length). For those interested in the use of alternative data to constrain the stochastic background of GWs, we refer to~\cite{Meerburg_2015, Cabass_2016}.

\subsection{Decoherence process}\label{subsec: Decoherence criterion}
In \cite{Zurek:1981xq, Joos1985TheEO, 1801.09949} it is shown how the non-unitary term in the evolution equation (\ref{eq:Lindblad equation for rho sys}) of the reduced density matrix of the system, that models the interaction with environmental degrees of freedom, leads to the dynamical suppression of its off-diagonal elements in the basis of the eigenstates of the interaction operator. This allows us to calculate the required interaction strength that leads to decoherence at the end of inflation (hence leading the quantum-to-classical transition of the primordial fluctuations). To quantify this, the parameter $\delta_{\boldsymbol{k}}$ is introduced for scalar perturbations, which characterizes the additional decrease in off-diagonal elements produced by the environment. This can be linked to the purity of the state\footnote{The purity of the state is defined as $\mathrm{Tr}(\hat{\rho}_{\boldsymbol{k}}^{2}) = \frac{1}{\sqrt{1+4\delta_{\boldsymbol{k}}}} \simeq 1 - 2\delta_{\boldsymbol{k}}$. This is valid at lowest order in the coupling parameter $(g^2)$.}, where $\delta_{\boldsymbol{k}} \ll 1$ means that the state remains pure and for $\delta_{\boldsymbol{k}}\gg 1$ the state is highly mixed (for more details see, e.g.,  \cite{Burgess_2008,1801.09949, 2024arXiv240312240B}). Therefore, successful decoherence is characterized by the condition $\delta_{\boldsymbol{k}}^s \gg 1$.
Additionally, in~\cite{1801.09949} a generic formula for this parameter is derived in the case of a scalar perturbation system, that we have explicitly checked to hold for our tensor case\footnote{Here, as a first approximation, we assume that the two polarization states are independent, i.e. we assume that the reduced density matrix is a product over the two polarization states, leading to the two separate quantities $\delta_{\boldsymbol{k}}^s$ for each polarization state.}
\begin{equation}
    \delta_{\boldsymbol{k}}^s(\eta) = \frac{1}{2} \int_{-\infty}^{\eta} S(\boldsymbol{k}, \eta') P_{vv}(\boldsymbol{k},\eta') \mathrm{d}\eta'.
\end{equation}
Therefore we apply it by inserting Eq. (\ref{eq:source function computed}). As this calculation is performed at linear order in $\gamma$, $P_{vv}$ is evaluated in the free theory. Plugging in our source function (\ref{eq:source function computed}) gives
\begin{equation}
    \delta_{\boldsymbol{k}}^s(\eta) = \frac{2\beta^2\sigma_{\gamma}}{9\sin^{2}{(\nu\pi)}} \bigg( \frac{k}{k_{\ast}} \bigg)^{p+1} \big[I_{1}(\nu, k, \eta) + I_{1}(-\nu, k, \eta) - 2I_{2}(\nu, k, \eta)\cos{(\nu\pi)}\big].
\end{equation}

\begin{figure}%
    \includegraphics[scale=0.51]{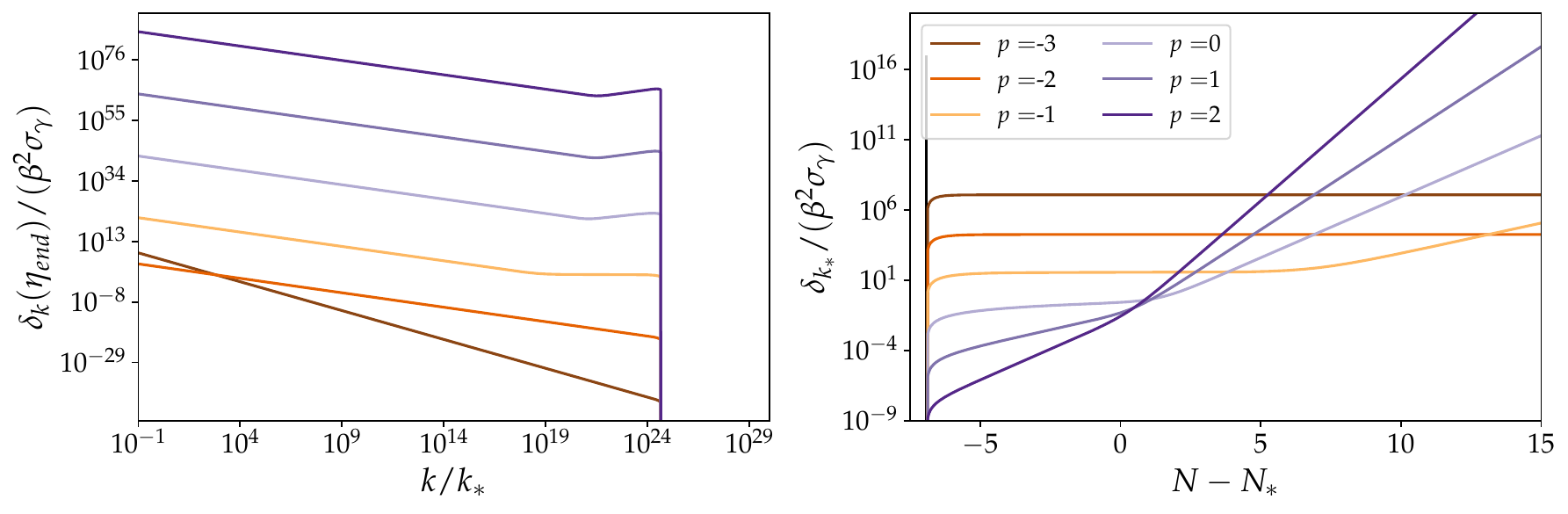}
    \caption{The parameter ($\delta_{\boldsymbol{k}}^s$) characterizing the decrease of the off-diagonal elements due to the interaction with the environment (rescaled by $\beta^2\sigma_{\gamma}$), for various $p$ scenarios. Left: the value of $\delta_{\boldsymbol{k}}^s$ at the end of inflation are shown depending on the scales, which are chosen to be inside the range of observational interest, from CMB to LVK scales. Right: the time evolution is shown, for CMB scales, in number of $e$-folds since the pivot scale exited the Hubble horizon. Assuming de Sitter inflation, the other model parameters are chosen to be $H_{\ast}l_E = 10^{-3}$, $\Delta N_{\ast} = N_{end} - N_{\ast} = 50$ and $N_T = N_{end} - N_{IR} = 10^{4}$.}%
    \label{fig:both deltak plots}%
\end{figure}

The evolution of $\delta_{\boldsymbol{k}}^s$ is shown in Figure \ref{fig:both deltak plots} for different values of $p$. On the left, we show the scale dependence of $\delta_{\boldsymbol{k}}^s$ when evaluated at the end of inflation. This illustrates the fact that for small scales exiting the horizon at a later time, their off-diagonal elements are less suppressed at the end of inflation. This seems to be consistent with the intuition that the smaller the scale, the less there is from horizon-crossing until the end of inflation, for that perturbation mode. Therefore, we expect the decoherence process to happen for a shorter period of time and thus be less effective.  On the right, we show the time evolution of $\delta_{\boldsymbol{k}}^s$ for the pivot scale. This clearly illustrates that higher values of $p$ have a much steeper increase over time. Looking at Eq. (\ref{eq: gamma dependence on a}) this makes sense, as a higher value of $p$ means the interaction strength increases more steeply over time. The stronger interaction strength means that over time the pivot scale will decohere rapidly and will have decohered more by the end of inflation.

The equation for $\delta_{\boldsymbol{k}}^s$ can be rewritten to calculate the minimum interaction strength required to achieve the complete classicalization of our perturbations on a given scale due to this interaction,before the end of inflation:
\begin{equation}\label{eq: lower lim sigma from delta k}
    \beta^2\sigma_{\gamma} \gg \frac{9 \sin^{2}(\nu\pi)}{2} \bigg( \frac{k}{k_{\ast}} \bigg)^{-p-1} \big[ I_{1}(\nu, k, \eta) + I_{1}(-\nu, k, \eta) - 2I_{2}(\nu, k, \eta)\cos{(\nu\pi)} \big]^{-1}.
\end{equation}
Assuming the CMB scales have become fully classical due to this interaction by the end of inflation and combing this with observational constraints from the CMB and LVK data, leads to the allowed region for $\delta_{\boldsymbol{k}}^s$ as shown by the blue contour in Figure \ref{fig:allowed region for sigma}. Let us note, that since the gravitational wave background from inflation has not been observed yet, it is not necessarily true that all the CMB scales have completely decohered. However, due to the amount of time spent outside the environmental correlation length scale, it is reasonable to assume that they have, in fact, decohered fully.

On the other hand, by exploiting Eq.(\ref{eq: lower lim sigma from delta k}) we can determine (for a given correlation length of the environment $l_{E}$, interaction strength $\beta^2\sigma_{\gamma}$ and time dependence $p$) on which scales decoherence would not have been completed by the end of inflation ($\delta_{\bf k} <1 $). This information can provide an indication of what scales to probe in order to possibly find some signatures of the (relic) quantumness of the inflationary gravitational waves. Given the recent direct detection of gravitational waves, there has been a renewed interest in looking for these quantum signatures, with various proposals both using interferometers and using table-top experiments \cite{ 2022Quant...6..879G, 2021LRR....24....4A, Parikh_2020, Parikh_2021, Parikh_2021_signatures, Lamine_2006,Kanno:2018cuk,Kanno:2019gqw} \footnote{It is also possible to test the quantum nature of inflationary perturbations with CMB data~\cite{PhysRevD.61.024024, Banerjee_2023, Martin:2015qta}, however this is will be very difficult. See also~\cite{Tejerina-Perez:2024opu} for a first attempt to look at the quantumness of the primordial perturbations via Large-Scale-Structure observables.}. If in the future we will have some firm observational probes or constraints on the quantum nature of primordial gravitational waves, then this in its turn will allow to put tighter constraints on the parameter space of the open quantum system under consideration. 

In fact, a crucial question we will address below is: given the parameter space allowed by imposing that CMB scales are fully decohered, what are the scales at which the tensor perturbations have not fully decohered? And more importantly, are future direct interferometers and proposed high-frequency experiments able to probe such frequency ranges and possibly detect these quantum signatures? 

\subsection{Quantum signatures}\label{subsec: quantum signatures}

\begin{figure}%
    \includegraphics[width=0.7\textwidth]{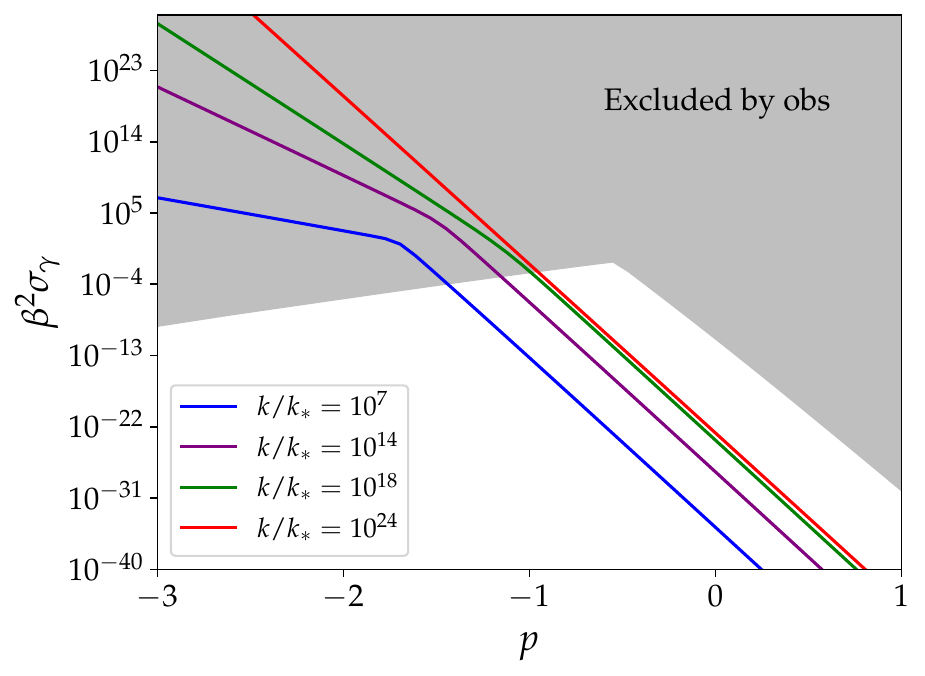}
    \caption{The maximum interaction strength, allowing for quantum signatures at specific scales, namely those probed by PTA ($k/k_{\ast}=10^{7}$), by space-based interferometers ($k/k_{\ast}=10^{14}$), by ground-based interferometers ($k/k_{\ast}=10^{18}$) and by possible future high frequency detectors ($k/k_{\ast}=10^{24}$). This is shown for various decoherence scenarios ($p$-values). For reference the maximum interaction strength allowed by observations is shown, assuming a Starobinsky model of inflation ($r_{\ast}=0.00461$). Assuming de Sitter inflation, the other model parameters are chosen to be $H_{\ast}l_E = 10^{-3}$, $\Delta N_{\ast} = N_{end} - N_{\ast} = 50$ and $N_T = N_{end} - N_{IR} = 10^{4}$.}%
    \label{fig: quantum signatures}%
\end{figure}

The evolution of $\delta_{\boldsymbol{k}}^s$ (Fig. \ref{fig:both deltak plots}) shows us that depending on the decoherence scenario and the interaction strength, not all scales need to have fully decohered due to this interaction, by the end of inflation. This naturally leads to the question: can gravitational waves from inflation still have some quantum signatures?\footnote{The criteria adopted here, to establish when relic quantum signatures might have survived, is a conservative one. The criterion is based on the determination of scales on which gravitational waves from inflation have not fully decohered. However, as shown in~\cite{Martin:2022kph,Martin2023}, there are cases where decoherence has been reached but quantum signatures (in the form of, e.g., quantum discord or violation of Bell inequalities) are large. As explained in~\cite{Martin:2022kph,Martin2023} this is due to a competition between quantum decoherence on side and quantum squeezing of the system on the other side: the larger the squeezing, the smaller the purity needs to be to erase quantum effects. A more extensive search for quantum signatures is left for future work.} In Figure \ref{fig: quantum signatures}, we show the maximum interaction strength for which specific scales have not fully decohered by the end of inflation. This is shown for different decoherence scenarios and for scales corresponding to those probed by PTA ($k/k_{\ast}\sim 10^{7}$), by space-based interferometers ($k/k_{\ast}\sim 10^{14}$), by ground-based interferometers ($k/k_{\ast}\sim 10^{18}$) and by possible future high frequency detectors ($k/k_{\ast}\sim 10^{24}$). It is clear from Fig. \ref{fig: quantum signatures} that for $p>-1$ scenarios, we need much more sensitive instruments to be able to probe these decoherence scenarios and look for possible quantum signatures. For $p \leq -1$, it is more difficult to see the effects of quantum decoherence, due to the shape of the power spectrum (see Fig. \ref{fig:Power spectrum}), but a much larger range $\beta^2\sigma_{\gamma}$ allows for leftover quantum signatures.

Let us note some caveats about possible relic quantum signatures. Firstly, here we only focus on our model of quantum decoherence, and we do not take the additional decoherence from other processes into account. As discussed in Section \ref{subsec: tensor perturbations} and Appendix \ref{sec:Daddi}, there are other decoherence processes that we expect to happen during inflation, making this a simplification. Secondly, we are considering $\delta_{\boldsymbol{k}}^s$ at the end of inflation, which means that these are quantum signatures that would be present at the end of inflation. However, let us note that since we are dealing with gravitational waves, it is well conceivable that once they reach a given state at the end of inflation, they remain so. This because, once reentered inside the horizon, they will essentially free-stream, being extremely weakly coupled to matter.

\subsection{Future detectors}\label{subsec: Future obs}
There are several future projects that focus on obtaining more GW data that can be used to further constrain the primordial GW power spectrum and the properties of GW. The constraints are on different frequency ranges and allow us to further constrain different decoherence scenarios. This can be achieved from direct detection with, for example, the space-based laser interferometer LISA \cite{LISA, Bartolo:2016ami, Barausse:2020rsu, Auclair_2023}, the Einstein Telescope \cite{Maggiore_2020, Branchesi_2023}, Cosmic Explorer \cite{reitze2019cosmic, evans2021horizon}, fourth generation gravitational wave detectors like BBO/DECIGO \cite{2011PhRvD..83d4011Y, Kawamura_2011, Corbin_2006} and possible ultra high-frequency detectors \cite{2021LRR....24....4A}. Additionally, this can be done using the Pulsar Timing Arrays method \cite{Verbiest_2021, 2023ApJ...951L...8A, 2023arXiv230616226A, 2023ApJ...951L...6R, InternationalPulsarTimingArray:2023mzf, Xu_2023} and future CMB experiments \cite{2023MNRAS.520.2405B, Ade_2019, 2020SPIE11443E..2FH, 2023arXiv231205194N, 2022ApJ...926...54A}. Specifically, future CMB experiments will be able to further constrain $r_{\ast}$. The current goal is to reach sensivities able to probe the Starobinsky model of inflation $r=0.00461$, shown in Figure \ref{fig:allowed region for sigma} as the dashed black line for reference. 

The benefit of ultra-high-frequency detectors \cite{2021LRR....24....4A}, is that they could be used to constrain the parameter $H_{\ast}l_{E}$, which we cannot do using current observational constraints. Indeed, as can be seen from Eqs. (\ref{eq:I1,I2 definition},\ref{eq: PT full}), the increase of the power spectrum is only on scales which have crossed out of the environmental correlation length ($-k\eta_E = (H_{\ast}l_E)^{-1}$). This leads to a small-scale cut off of the decoherence-induced increase of the power spectrum, which (for the values of the parameters considered here) is on scales much smaller than our current detectors. With future high-frequency detectors, we could reach these very small scales and constrain $H_{\ast}l_E$.

We now study which decoherence scenarios could be observed by two future detectors, namely LISA and ET. Due to the frequency range of these detectors, we focus only on the scenarios with $p>-1$, since these show an increase in the power spectrum due to decoherence on small scales.\footnote{We assume the constraints currently set by both the CMB and LVK hold.} Using the predicted sensitivity of these detectors together with the approach layed out in Section~\ref{subsec: Observational constraints}.\footnote{We are using the exact same approach but now $\mathcal{P}_{T|\mathrm{obs}}(k)$ is taken to be the power spectrum that can be observed by a future detector. Then requiring  $\mathcal{P}_{T}(k) > \mathcal{P}_{T|\mathrm{obs}}(k)$ allows us to probe which decoherence scenarios would be visible with these instruments. Note that this is an estimation depending on the final designs of both LISA and ET and therefore serves only as an indication.}

\begin{figure}%
    \includegraphics[width=\textwidth]{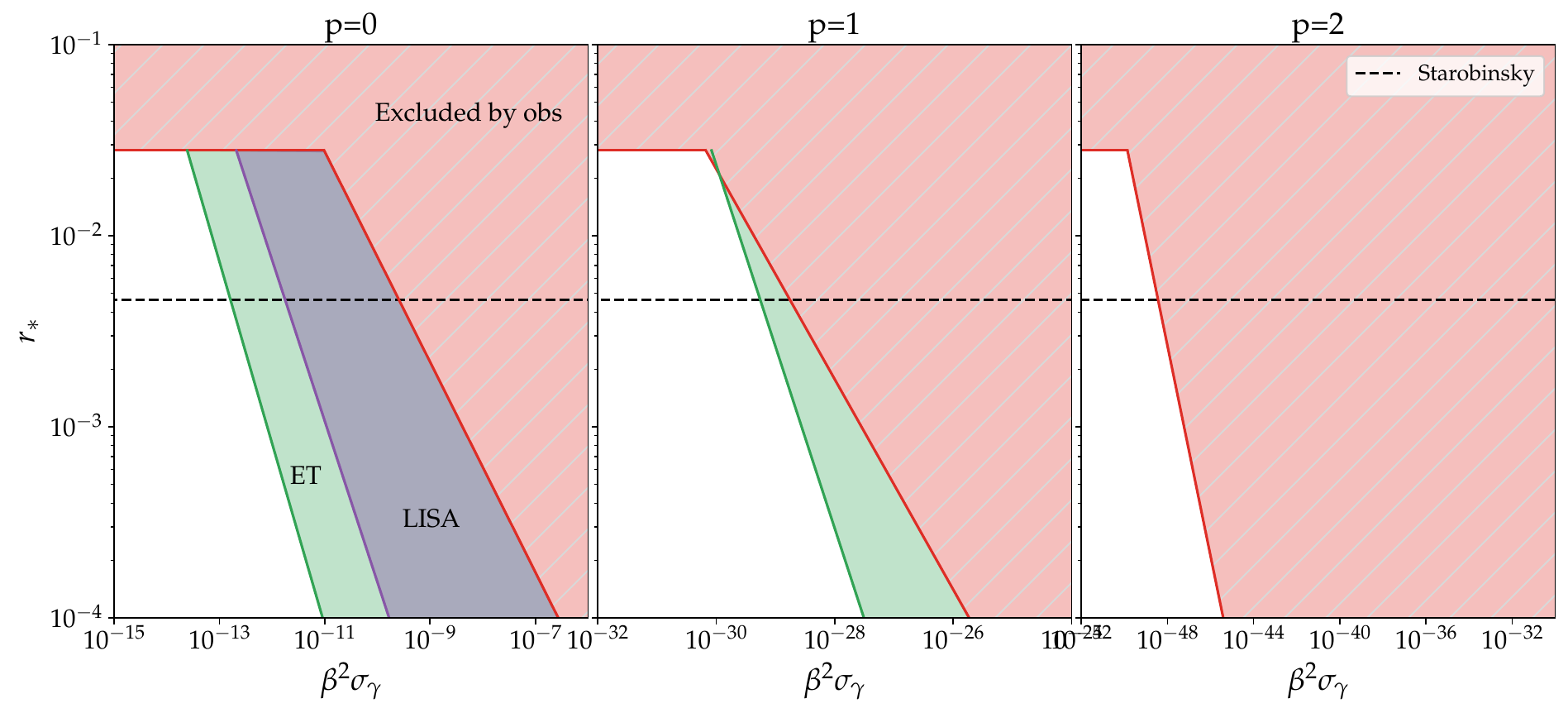}
    \caption{In red the values of $\beta^2\sigma_{\gamma}$ excluded by observations done at scales ranging from CMB to LVK, shown for a range of $r_{\ast}$ indicating the amplitude of the spectrum at $k_{\ast} = 0.05 \rm{Mpc^{-1}}$ and for three different values of $p$, each corresponding to the $p>-1$ scenario. The green (purple) region indicates the scenarios would be seen with the future detector ET (LISA). For reference the Starobinsky model of inflation ($r_{\ast}=0.00461$) is shown. Assuming de Sitter inflation, the other model parameters are chosen to be $H_{\ast}l_E = 10^{-3}$, $\Delta N_{\ast} = N_{end} - N_{\ast} = 50$ and $N_T = N_{end} - N_{IR} = 10^{4}$. }%
    \label{fig:future obs}%
\end{figure}

For LISA, projected sensitivities \cite{2016JCAP...12..026B} have been used to study which decoherence scenarios could be investigated. Assuming a power law spectrum (\ref{eq:power law observations}), LISA will be able to observe stochastic background GWs for a specific region of $(r,n_T)$, see Figure 3 of \cite{2016JCAP...12..026B}. These observations would correspond to specific decoherence scenarios, shown in Figure \ref{fig:future obs}. Unfortunately, due to the frequency range and expected sensitivity of LISA, we will only be able to observe/exclude very specific decoherence scenarios. Therefore, our main focus will be on ET, for which observable scenarios can be computed using the power-law integrated sensitivity estimated for ET \cite{Branchesi_2023}. Expressing the GW energy density fraction today in terms of the scalar ratio $r$ and the spectral index $n_{T}$,\footnote{Using 
\begin{equation*}
    \Omega_{\rm GW} (f) = \frac{3}{128} \Omega_{\rm rad} r A_{s}\bigg( \frac{f}{f_{\ast}} \bigg)^{n_{T}} \bigg[ \frac{1}{2} \bigg( \frac{f_{\rm eq}}{f} \bigg)^2 + \frac{16}{9} \bigg],
\end{equation*}
where $f_{\ast} = k_{\ast}/(2\pi a_{0})$ is the pivot frequency, $f_{\rm eq} = H_0\Omega_{\rm mat}/(\pi \sqrt{2\Omega_{\rm rad}})$ is the frequency entering the horizon at matter-radiation equality, and $\Omega_{\rm rad}$  is the radiation energy density.} \cite{2018CQGra..35p3001C}, allows us to employ the same method shown above. This shows that we will be able to probe the decoherence scenarios shown in Figure \ref{fig:future obs} with ET. The same analysis has been applied to the model presented in \cite{2211.07598}, for which the results are shown in Appendix \ref{sec:Daddi}.

It is clear from Figure \ref{fig:future obs} that for $p=0$ the constraints on $\beta^2\sigma_{\gamma}$ can increase by more than three orders of magnitude, just using LISA, and even more when using ET. However, unfortunately, for higher values of $p$ we cannot use LISA to further constrain the decoherence scenario, as increasing $p$ leads to an increase in the slope of the power spectrum, which would then have already been observed with LVK. For $p=1$ ET could still increase our constraints but for higher values of $p$ we encounter the same problem as for LISA. This is another good motivation to look at other future detectors such as DECIGO/BBO\cite{2011PhRvD..83d4011Y, Kawamura_2011, Corbin_2006}. Not only because we will be able to constrain $p>-1$ scenarios much further, due to the power spectrum increase on small scales, but also, because their high sensitivities allows us to further constrain $r_{\ast}$, therefore also putting further constraints on $p\leq -1$ scenarios. Similarly, future CMB experiments will be able to further constrain $r_{\ast}$. The goal is to get a sensitivity of $r_{\ast}=0.00461$, which would allow us to constrain the entire parameter space above the dashed black line shown in Fig. \ref{fig:allowed region for sigma}. 

Unfortunately, as explained above, PTA observes on frequencies and with a sensitivity that will not allow us to further constrain the quantum decoherence models considered here. This is because any scenario that could be probed with PTA, would also need to be seen by either CMB or LVK. Still, PTA still provides important information about the stochastic GW background.



\section{Conclusions}\label{sec: Conclusion}

In this work, we studied the effect of the quantum decoherence of inflationary gravitational waves on their power spectrum, adopting the approach of open quantum systems used in \cite{1801.09949} and first applied to tensor perturbations in \cite{2211.07598}. The system is given by tensor perturbations, and the environment is another field present during inflation, which (within an open effective field theory approach) we do not specify but we assume satisfies certain conditions (like the assumption that the environmental correlation length is much smaller than the Hubble radius). In particular, we extend on previous works by using the quadratic system interaction operator containing the partial spatial derivatives of tensor perturbations, which naturally appear within GR. 

Using the Lindblad equation allows us to describe the dynamics of the system, giving the time evolution of the two-point correlators of the tensor perturbations and its momentum conjugate. This system of equations can be combined into a third-order differential equation for the GW power spectrum, including a source term due to decoherence. Calculating the source term explicitly allows us to obtain the expression for the full GW power spectrum, shown in Figure \ref{fig:Power spectrum}. Looking at the GW power spectrum, it is clear that there are three possible scenarios, depending on the time dependence of the interaction strength (described by the parameter $p$). The first option is that the power spectrum increases as a power law on large scales (small $k$) due to decoherence and remains unchanged at small scales. Second, the power spectrum increases on small scales (large $k$) and remains unchanged on large scales. Lastly, the power spectrum has a scale-independent change. 

The main parameter focused on in this work is the interaction strength ($\beta^2\sigma_{\gamma}$), whose value changes the power spectrum as shown in Figure \ref{fig:different sigma}. This interaction strength can be constrained in two ways. First, using observational constraints on the spectral tilt and amplitude of the GW power spectrum \cite{Galloni_2023}. This gives an upper bound on the interaction strength, depending on its time dependence $p$ and the amplitude of the power spectrum without decoherence effects. Secondly, an indication of a lower bound on the interaction strength can be obtained by using the constraints set by the decoherence process itself. In order for decoherence to explain the quantum-to-classical transition, it is important that decoherence lasts long enough for the perturbations to become classical, thus providing a lower limit on the interaction strength. Let us note, this is dependent on the decoherence scenario (see Fig. \ref{fig:both deltak plots}) and assumes the classicalization of the perturbations is only due to this type of decoherence. Combining both observational constraints and the requirement that primordial tensor perturbations are fully decohered on CMB scales allows us to constrain the parameter space as shown in Figure \ref{fig:allowed region for sigma}.

In addition to this, we study which decoherence scenarios could be probed using future GW detectors like LISA and ET. It is shown that especially with ET, but also with LISA, specific decoherence scenarios (corresponding to a range of $p$ and $\beta^2\sigma_{\gamma}$ values) could be observed and the interaction strength can be tighter constrained. However, for higher values of $p$ the slope of the power spectrum is steeper, so that it quickly approaches the regime already probed by LVK, which therefore is already putting tight constraints on these cases.
Similarly for values $p \leq - 1$, any scenario that could be probed with future GW detectors would already have been ruled out by the CMB data due to the shape of the power spectrum. This means that in order to constrain the parameter space even further we need to use future detectors with an even higher sensitivity, like DECIGO/BBO \cite{2011PhRvD..83d4011Y}, and we need to determine the amplitude of the GW power spectrum (without decoherence effects), possibly using CMB observations from Simons Observatory \cite{Ade_2019}, CMB-S4 \cite{2022ApJ...926...54A} or LiteBIRD \cite{LiteBIRD:2022cnt}.

We have also investigated under which conditions and, most importantly, on which scales the decoherence process might not be completed. This leaves an open window to test the quantumness of inflationary tensor fluctuations (see Fig.~\ref{fig: quantum signatures}). Within the scenarios considered here, such relic quantumness of primordial GWs will be hard to be tested at future interferometers like ET and LISA, if we require the corresponding power-spectrum amplitude to be measured in the first instance. Nevertheless we believe this computation and this investigation is valuable as a pathway to future studies for different decoherence scenarios.



\begin{acknowledgments}
The authors would like to thank Sabino Matarrese, Francescopaolo Lopez, Vincent Vennin, Flaminia Giacomini, Daniele Bertacca and Valentina Danieli for many useful and stimulating discussions.  
We thank Vincent Vennin and Jerome Martin also for some clarifications.
We would also like to thank Angelo Ricciardone, Ilaria Caporali, Sarah Libanore and Federico Semenzato for their useful discussions about gravitational wave detections, and Marco Peloso and Nicola Bellomo for usefull feedbacks and suggestions. 
A special thanks to Aoumeur Daddi Hammou for his help in the very first phase of the work and for valuable interactions. 
We acknowledge financial support from the INFN InDark initiative. N.B. acknowledges financial support from the COSMOS network (www.cosmosnet.it) through the ASI (Italian Space Agency) Grants 2016-24-H.0, 2016-24-H.1-2018 and 2020-9-HH.0. 
N.B. acknowledges support by the MUR PRIN2022 Project “BROWSEPOL: Beyond standaRd mOdel With coSmic microwavE back- ground POLarization”-2022EJNZ53  nanced by the European Union - Next Generation EU. 
This work is supported in part by the MUR Departments of Excellence grant "Quantum Frontiers".
\end{acknowledgments}

\appendix

\section{Computation of the source function}\label{sec:computation source function}

Looking at Eq. (\ref{eq: source function}) we start by working out the integrals, similarly to Appendix D.1 of \cite{1801.09949}. In the following, we show this computation. 
The integral $I$ we need to compute is given by
\begin{equation}
    I = \int \mathrm{d}^3\boldsymbol{k'} \Bigl( (\boldsymbol{k}\cdot \boldsymbol{k'})^2 + k^4 + 2k^2(\boldsymbol{k}\cdot\boldsymbol{k'})\Bigr) \Tilde{C}_R (|\boldsymbol{k'}|) P_{vv}(|\boldsymbol{k}+\boldsymbol{k'}|).
\end{equation}
To simplify the computation, we perform a change of integration variable $\boldsymbol{k'} = \boldsymbol{p}-\boldsymbol{k}$ and negate the prefactors for simplicity. We choose $\theta$ to be the angle between $\boldsymbol{p}$ and $\boldsymbol{k}$. Therefore, the term with the dot product becomes
\begin{equation}
    (\boldsymbol{k}\cdot \boldsymbol{k'})^2 + k^4 + 2k^2(\boldsymbol{k}\cdot\boldsymbol{k'}) = (\boldsymbol{k}\cdot (\boldsymbol{p}-\boldsymbol{k}))^2 + k^4 + 2k^2(\boldsymbol{k}\cdot(\boldsymbol{p}-\boldsymbol{k})) = k^2p^2\cos{\theta}^2.
\end{equation}
Plugging this back into the integral and changing to spherical coordinates, where $\theta$ is the polar angle, one has
\begin{equation}
    I = 2\pi \int_0^{\infty} \mathrm{d}p p^2 \int_0^{\pi} \mathrm{d}\theta \sin{\theta} k^2p^2\cos{\theta}^2\Tilde{C}_R (\sqrt{k^2+p^2-2kp\cos{\theta}}) P_{vv}(p),
\end{equation}
where the integral over the azimuth angle has been performed. Using the change of variable $z=k^2+p^2-2kp\cos{\theta}$, one obtains
\begin{equation}\label{eq:I in angular coordinates}
    I = \frac{\pi}{4k} \int_0^{\infty} \mathrm{d}p p  P_{vv}(p) \int_{(k-p)^2}^{(k+p)^2} \mathrm{d}z (k^2+p^2-z)^2 \Tilde{C}_R (\sqrt{z}).
\end{equation}
This allows us to perform one-dimensional integrals only.

We perform the second integral using the ansatz shown in Eq. (\ref{eq: C tilde with sin cos}) for the environment correlator. Doing this integral exactly, the integral becomes
\begin{equation}\label{eq:I first step}
\begin{aligned}
    I &= \frac{\pi \Bar{C_R}}{4k}\sqrt{\frac{2}{\pi}}\int_{0}^{\infty} \mathrm{d}p p P_{vv}(p) \Biggl[ \Bigg( \frac{16a^3}{l_E^3} - \frac{16apk}{l_E} \Biggr)\cos{\biggl( \frac{l_E |k+p|}{a} \biggr)} \\
    &-\frac{2}{l_E^2|k+p|}\bigl(4p^2k^2l_E^2 -8a^2(k+p)^2\bigr)\sin{\biggl( \frac{l_E |k+p|}{a} \biggr)} +\Bigg( \frac{-16a^3}{l_E^3} - \frac{16apk}{l_E} \Biggr) \cos{\biggl( \frac{l_E |k-p|}{a} \biggr)} \\
    &+\frac{2}{l_E^2|k-p|}\bigl(4p^2k^2l_E^2 -8a^2(k-p)^2\bigr)\sin{\biggl( \frac{l_E |k-p|}{a} \biggr)}\Biggr].
\end{aligned}
\end{equation}

Next, we insert the power spectrum into the above equation and perform the integral. For simplicity, we neglect the slow-roll corrections and use the piecewise approximation for the power spectrum. We split the integral into sub-Hubble scales (UV part) with $-p\eta > 1$, for which we have $P_{vv}(p) = (2p)^{-1}$, and super-Hubble scales (IR part) with $-p\eta <1$, for which we have $P_{vv}(p) = (2p)^{-1}(-p\eta)^{-2}$. This leads to $I = I_{IR} + I_{UV}$, with
\begin{equation}\label{eq:IR integral}
\begin{aligned}
    I_{IR} &= \frac{\pi \Bar{C_R}}{8k\eta^2}\sqrt{\frac{2}{\pi}}  \int_{0}^{-1/\eta} \frac{\mathrm{d}p}{p^2}   \Biggl[ \Bigg( \frac{16a^3}{l_E^3} - \frac{16apk}{l_E} \Biggr)\cos{\biggl( \frac{l_E |k+p|}{a} \biggr)} \\
    &-\frac{2}{l_E^2|k+p|}\bigl(4p^2k^2l_E^2 -8a^2(k+p)^2\bigr)\sin{\biggl( \frac{l_E |k+p|}{a} \biggr)} + \Bigg( \frac{-16a^3}{l_E^3} - \frac{16apk}{l_E} \Biggr)\cos{\biggl( \frac{l_E |k-p|}{a} \biggr)} \\
    &+\frac{2}{l_E^2|k-p|}\bigl(4p^2k^2l_E^2 -8a^2(k-p)^2\bigr)\sin{\biggl( \frac{l_E |k-p|}{a} \biggr)}\Biggr],
\end{aligned}
\end{equation}
\begin{equation}\label{eq:UV integral}
\begin{aligned}
    I_{UV} &= \frac{\pi \Bar{C_R}}{8k}\sqrt{\frac{2}{\pi}} \int_{-1/\eta}^{\infty} \mathrm{d}p \Biggl[ \Bigg( \frac{16a^3}{l_E^3} - \frac{16apk}{l_E} \Biggr)\cos{\biggl( \frac{l_E |k+p|}{a} \biggr)} \\
    &-\frac{2}{l_E^2|k+p|}\bigl(4p^2k^2l_E^2 -8a^2(k+p)^2\bigr)\sin{\biggl( \frac{l_E |k+p|}{a} \biggr)} + \Bigg( \frac{-16a^3}{l_E^3} - \frac{16apk}{l_E} \Biggr)\cos{\biggl( \frac{l_E |k-p|}{a} \biggr)} \\
    &+\frac{2}{l_E^2|k-p|}\bigl(4p^2k^2l_E^2 -8a^2(k-p)^2\bigr)\sin{\biggl( \frac{l_E |k-p|}{a} \biggr)}\Biggr].
\end{aligned}
\end{equation}

The UV integral (\ref{eq:UV integral}) is usually removed by adiabatic subtraction \cite{T_S_Bunch_1980, Markkanen_2018}, therefor we do not consider it and instead focus on the IR integral (\ref{eq:IR integral}). 

The IR integral does not converge so we introduce an IR cut off and replace the lower bound of the integral $0$ with $-1/\eta_{IR}$, which can be seen as the comoving mode that corresponds to the Hubble radius at the onset of inflation. The integral can be performed explicitly, and to simplify we define two parameters $K \equiv kl_E/a$ and $y \equiv l_Ep/a$. The integral becomes
\begin{equation}\label{eq:I IR with I(p)}
    I_{IR} (\eta) = \frac{\pi \Bar{C_R}}{8k\eta^2}\sqrt{\frac{2}{\pi}} \Bigg[ \mathcal{I}\biggl( \frac{-l_E}{\eta a}\biggr) -\mathcal{I}\biggl( \frac{-l_E}{\eta_{IR}a}\biggr) \Bigg],
\end{equation}
where we have introduced the function $\mathcal{I}$ defined by
\begin{equation}
    \mathcal{I}(y) = \frac{32a}{l_E^2y} \sin{y}\Big[ a\sin{K} - aK\cos{K}  \Big] -\frac{8a^2K^2}{l_E^2} \Big[ \mathrm{Si}(K+y) + \mathrm{Si}(K-y) \Big]
\end{equation}
We assume that the correlation length of the environment $l_E$ is much smaller than the Hubble radius $H^{-1}$, which is true if $l_E \sim t_c$ since the derivation of the Lindblad equation requires $t_c \ll H^{-1}$, see the Appendix A of \cite{1801.09949}. This amounts to $l_E/a\eta \ll 1$, which also implies that $l_E/a\eta_{IR} \ll 1$ since, by definition, $\eta > \eta_{IR}$ during inflation. As a consequence, we need to study the limit $y \to 0$, which gives
\begin{equation}
    \mathcal{I}(y) \simeq \frac{32a^2}{l_E^2} \Big[ \sin{K} - K\cos{K}  \Big] -\frac{16a^2K^2}{l_E^2}\mathrm{Si}(K) + \frac{8a^2y^2}{3l_E^2} \Big[ \sin{K} - K\cos{K}  \Big].
\end{equation}
Using the ansatz shown in Eq. (\ref{eq: C tilde with sin cos}) for the environment correlator the integral can be written as
\begin{equation}
    I_{IR} = \frac{\pi k^2\Tilde{C}_R(k)}{3\eta^2} \Bigl( \frac{1}{\eta^2} -\frac{1}{\eta_{IR}^2}\Bigr). 
\end{equation}
As the UV integral can be subtracted adiabatically, we can take $I\simeq I_{IR}$. This leads to the source function
\begin{equation}\label{eq:source function final}
    S(k,\eta) = \frac{8\gamma\beta^2}{(2\pi)^{3/2}}I =   \frac{2k^2\beta^2\gamma\Tilde{C}_R(k)}{3\eta^2} \sqrt{\frac{2}{\pi}} \Bigl( \frac{1}{\eta^2} -\frac{1}{\eta_{IR}^2}\Bigr).
\end{equation}



\section{Another contribution}\label{sec:Daddi}

In \cite{2211.07598} a specific model for the quantum decoherence of tensor perturbations is presented. The system operator used is 
\begin{equation}\label{eq:Daddi system operator}
    A_{D} (\eta, \boldsymbol{x}) = h_{ij}(\eta, \boldsymbol{x})h^{ij}(\eta, \boldsymbol{x}),
\end{equation}
where the subscript $D$ will be used to indicate this model.
This operator shows that the difference between the model presented in our work and the model presented in \cite{2211.07598} is the spatial partial derivatives. However, through partial integration of the Lagrangian, it can be shown that both of these terms can be present. Therefore, we expand on previous work done by studying how this model can be constrained using observations. Studying how exactly these two models and possible others need to be combined is left for future work. 

\begin{figure}%
    \includegraphics[width=\textwidth]{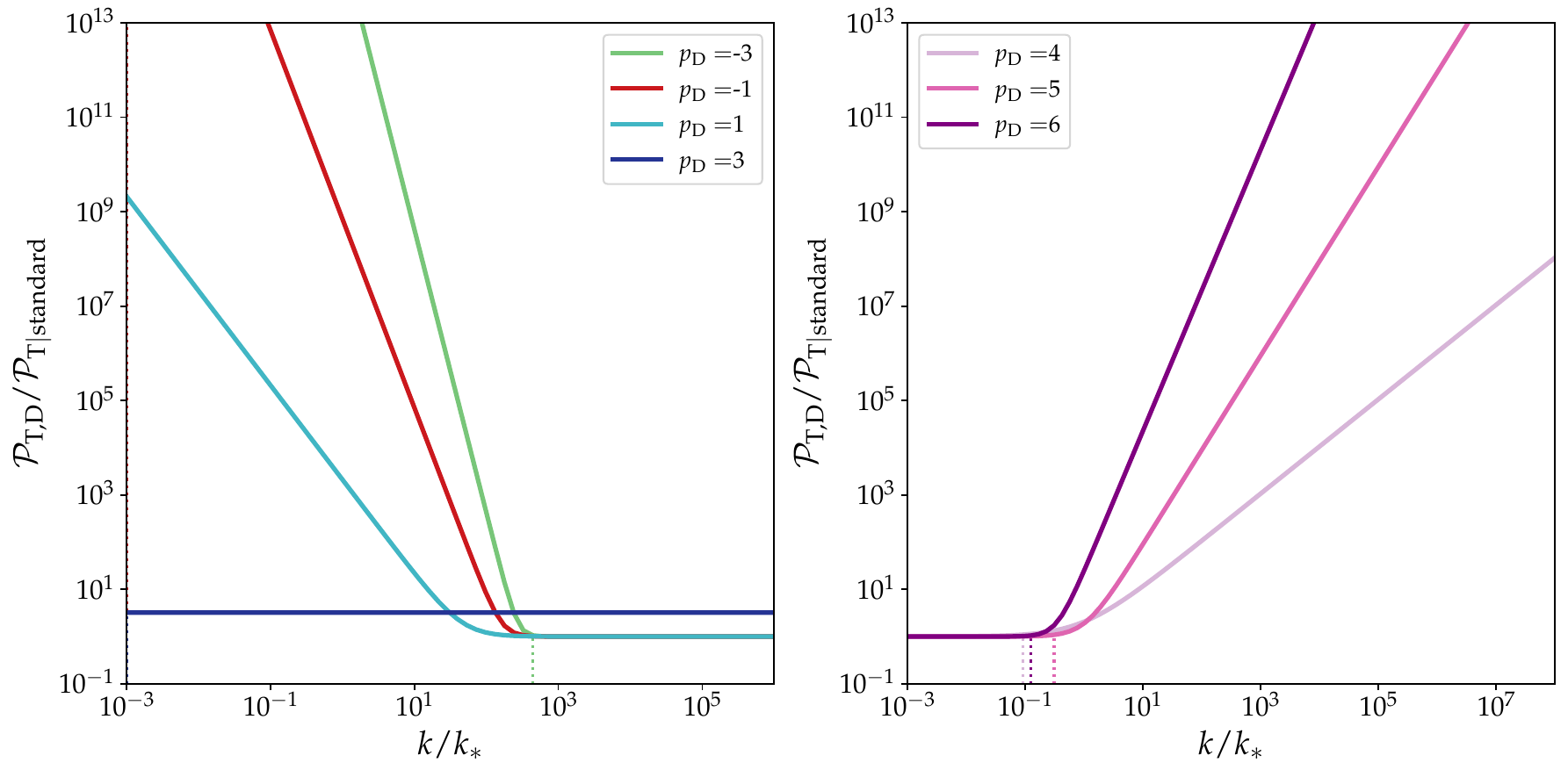}
    \caption{The GW power spectrum at the end of inflation, including the effects of decoherence. The colors represent different values of $p_{D}$. The dotted lines correspond to the transition scale $k_t$, defined by $\Delta P_{k_t}=1$. Assuming de Sitter inflation, the other model parameters are chosen to be $H_{\ast}l_E = 10^{-3}$, $\beta^2\sigma_{\gamma} = 10^{-3}$, $\Delta N_{\ast} = N_{end} - N_{\ast} = 50$ and $N_T = N_{end} - N_{IR} = 10^{4}$. }%
    \label{fig:Daddi Power spectrum}%
\end{figure}

The full expression for the power spectrum is given by
\begin{equation}\label{eq: Daddi PT full}
\begin{aligned}
    \mathcal{P}_{T,D}(k) = \mathcal{P}_{T|\mathrm{stan}} \Bigg( 1 + \frac{4 \beta_{D}^2 \sigma_{\gamma,D}}{3 \sin^2(\nu\pi)} \bigg( \frac{k}{k_{\ast}} \bigg)^{p_{D}-3} \frac{JI_{D}(k, \eta, \nu)}{[J_{\nu}^2(-k\eta) + Y_{\nu}^2(-k\eta)]} \Bigg).
\end{aligned}
\end{equation}
with $\beta_{D} = 2\xi_{D}/M_{pl}^2$, and we use 
\begin{equation}
    JI_{D}(k, \eta, \nu) =  J_{-\nu}^2(-k\eta)I_3(\nu, k, \eta) -2J_{-\nu}(-k\eta)J_{\nu}(-k\eta)I_4(\nu, k, \eta) + J_{\nu}^2(-k\eta)I_3(\nu, k, \eta) ,
\end{equation}
and the integrals $I_3$ and $I_4$ are defined by
\begin{equation}\label{eq:I3,I4 definition}
\begin{aligned}
    &I_3(\nu, k, \eta) \equiv \int_{-k\eta}^{(H_{\ast}l_{E,D})^{-1}} \mathrm{d}z z^{2-p_{D}} \ln{\bigg( \frac{-z}{k\eta_{IR}} \bigg)} J_{\nu}^2(z), \\
    &I_4(\nu, k, \eta) \equiv \int_{-k\eta}^{(H_{\ast}l_{E,D})^{-1}} \mathrm{d}z z^{2-p_{D}} \ln{\bigg( \frac{-z}{k\eta_{IR}} \bigg)} J_{\nu}(z)J_{-\nu}(z).
\end{aligned}
\end{equation}
This shows there is a $k^{4}$ shift with respect to our model. This is also apparent when looking at the power spectrum, shown in Figure \ref{fig:Daddi Power spectrum}. The change in power spectrum is now scale-invariant for $p_{D}=3$, and for values below (above) the change is on large (small) scales. 

\begin{figure}%
    \includegraphics[width=\textwidth]{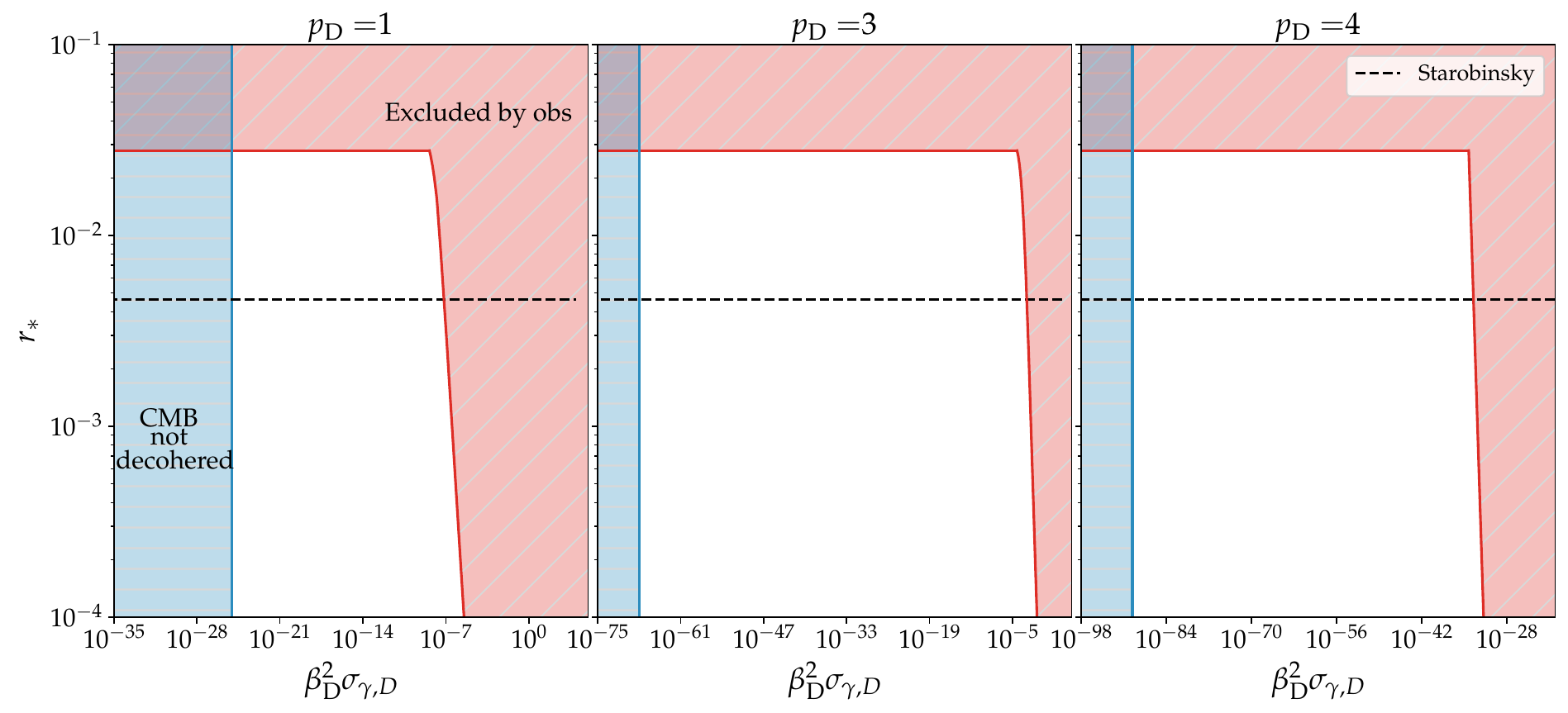}
    \caption{The values of $\beta_{D}^2\sigma_{\gamma,D}$ excluded by observations done at scales ranging from LVK, up to CMB, shown in red for a range of $r_{\ast}$ indicating the amplitude of the spectrum at $k_{\ast}$ and for three different values of $p$, each corresponding to a different scenario ($p_{D}<3, p_{D}=3, p_{D}>3$). The blue region indicates the scenarios for which the CMB scales have completely decohered. For reference the Starobinsky model of inflation ($r_{\ast}=0.00461$) is shown. Assuming de Sitter inflation, the other model parameters are chosen to be $H_{\ast}l_E = 10^{-3}$, $\Delta N_{\ast} = N_{end} - N_{\ast} = 50$ and $N_T = N_{end} - N_{IR} = 10^{4}$.}%
    \label{fig:Daddi allowed region for sigma}%
\end{figure}

For this model the decoherence criterion is given by
\begin{equation}
    \delta_{\boldsymbol{k}, D}^s = \frac{\beta_{tm}^2\sigma_{\gamma,D}}{3\sin^{2}{(\nu\pi)}} \bigg( \frac{k}{k_{\ast}} \bigg)^{p_{D}-3} \big[I_{3}(\nu, k, \eta) + I_{3}(-\nu, k, \eta) - 2I_{4}(\nu, k, \eta)\cos{(\nu\pi)}\big].
\end{equation}
Combining this with the observational constraints set by \cite{Galloni_2023}, in the same way as done in Section \ref{subsec: Observational constraints} allow us to find the allowed range for the interaction strength of this model, as shown in Figure \ref{fig:Daddi allowed region for sigma}.

\begin{figure}%
    \includegraphics[width=\textwidth]{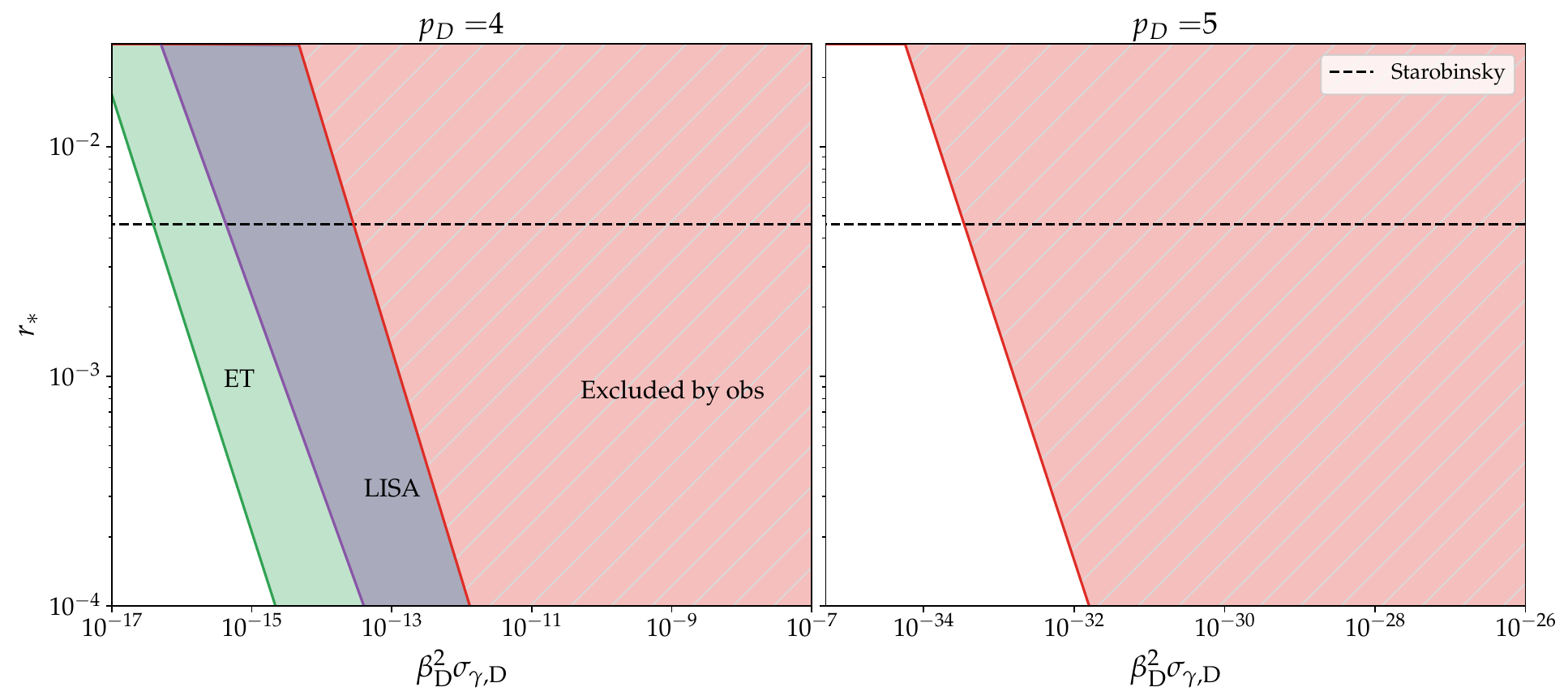}
    \caption{In red the values of $\beta_{D}^2\sigma_{\gamma,D}$ excluded by observations done at scales ranging from LVK, up to CMB, shown for a range of $r_{\ast}$ indicating the amplitude of the spectrum at $k_{\ast}$ and for two different values of $p$, each corresponding to the $p_{D}>3$ scenario. The light green (purple) region indicates the scenarios would be seen with the future detector ET (LISA). For reference the Starobinsky model of inflation ($r_{\ast}=0.00461$) is shown. Assuming de Sitter inflation, the other model parameters are chosen to be $H_{\ast}l_E = 10^{-3}$, $\Delta N_{\ast} = N_{end} - N_{\ast} = 50$ and $N_T = N_{end} - N_{IR} = 10^{4}$.}%
    \label{fig:Daddi future obs}%
\end{figure}

Analogue to Section \ref{subsec: Future obs}, we can see which decoherence scenarios could be probed with future detectors like LISA and ET. For this model we focus on scenarios with $p_{D}>3$, due to the shape of the power spectrum. The regions that could be probed are shown in Figure \ref{fig:Daddi future obs}. It is evident that for both our and this model, ET and LISA can further constrain the interaction strength in the same way. Therefore it is important for future work to focus on the interplay between the different decoherence models. 

\section{Polarization of gravitational waves}
\label{sec:polarization}
In this Appendix we provide the extension of the hierarchy of equations 
written in~(\ref{eq: first form two-point correlator functions}) which were 
used to compute  the power-spectrum of the gravitational waves. Here we consider also the two-point correlation functions between different polarization states for the tensor fluctuations.

Now we have the two-point correlators of the form $\langle \hat{O} \rangle = \langle \hat{O}_{\boldsymbol{k}_1}\hat{O}_{\boldsymbol{k}_2} \rangle$ with $\hat{O}_{\boldsymbol{k}_i} = \hat{v}_{\boldsymbol{k}_i}$ or $\hat{p}_{\boldsymbol{k}_i}$. Furthermore, we use the relation $[\hat{v}_{\boldsymbol{p}}^s, \hat{p}_{\boldsymbol{k}}^{\lambda}] = i\delta^{s\lambda}\delta^{(3)}(\boldsymbol{p}+\boldsymbol{k})$. The full computations and results for these correlators are shown below. Starting with the $\langle v_{\boldsymbol{k}_1}^s v_{\boldsymbol{k}_2}^{s'} \rangle$ one we get
\begin{equation}\label{eq:comp 1st observable}
\begin{aligned}
    &\frac{\mathrm{d}\langle v_{\boldsymbol{k}_1}^s v_{\boldsymbol{k}_2}^{s'} \rangle}{\mathrm{d}\eta} =  -i \langle [v_{\boldsymbol{k}_1}^sv_{\boldsymbol{k}_2}^{s'}, H_v] \rangle = \frac{-i}{2}\sum_{\lambda} \int \mathrm{d}^3 \boldsymbol{q} \langle [v_{\boldsymbol{k}_1}^sv_{\boldsymbol{k}_2}^{s'},p_{\boldsymbol{q}}^{\lambda}p_{-\boldsymbol{q}}^{\lambda}] \rangle \\
    &= \frac{-i}{2}\sum_{\lambda} \int \mathrm{d}^3 \boldsymbol{q} \langle v_{\boldsymbol{k}_1}^s[v_{\boldsymbol{k}_2}^{s'},p_{\boldsymbol{q}}^{\lambda}]p_{-\boldsymbol{q}}^{\lambda}+ [v_{\boldsymbol{k}_1}^s,p_{\boldsymbol{q}}^{\lambda}]v_{\boldsymbol{k}_2}^{s'}p_{-\boldsymbol{q}}^{\lambda} + p_{\boldsymbol{q}}^{\lambda}[v_{\boldsymbol{k}_1}^s,p_{-\boldsymbol{q}}^{\lambda}]v_{\boldsymbol{k}_2}^{s'} + p_{\boldsymbol{q}}^{\lambda}v_{\boldsymbol{k}_1}^s[v_{\boldsymbol{k}_2}^{s'},p_{-\boldsymbol{q}}^{\lambda}] \rangle \\
    &= \frac{-i}{2}\sum_{\lambda} \int \mathrm{d}^3 \boldsymbol{q} \langle i\delta^{(3)} v_{\boldsymbol{k}_1}^s(\boldsymbol{k}_2+\boldsymbol{q})\delta^{s'\lambda}p_{-\boldsymbol{q}}^{\lambda}+ i\delta^{(3)}(\boldsymbol{k}_1+\boldsymbol{q})\delta^{s\lambda}v_{\boldsymbol{k}_2}^{s'} p_{-\boldsymbol{q}}^{\lambda} + p_{\boldsymbol{q}}^{\lambda}i\delta^{(3)}(\boldsymbol{k}_1-\boldsymbol{q})\delta^{s\lambda}v_{\boldsymbol{k}_2}^{s'} \\
    &+ p_{\boldsymbol{q}}^{\lambda}v_{\boldsymbol{k}_1}^si\delta^{(3)}(\boldsymbol{k}_2-\boldsymbol{q})\delta^{s'\lambda} \rangle
    = \frac{\langle v_{\boldsymbol{k}_1}^s p_{\boldsymbol{k}_2}^{s'} + v_{\boldsymbol{k}_2}^{s'} p_{\boldsymbol{k}_1}^{s} + p_{\boldsymbol{k}_1}^s v_{\boldsymbol{k}_2}^{s'} + p_{\boldsymbol{k}_2}^{s'} v_{\boldsymbol{k}_1}^{s} \rangle }{2} = \langle v_{\boldsymbol{k}_1}^sp_{\boldsymbol{k}_2}^{s'} + p_{\boldsymbol{k}_1}^sv_{\boldsymbol{k}_2}^{s'} \rangle.
\end{aligned}
\end{equation}
Second, we look at the two correlators that contain both $ v_{\boldsymbol{k}_1}^s$ and $p_{\boldsymbol{k}_2}^{s'}$. 
\begin{equation}\label{eq:comp 2nd observable}
\begin{aligned}
    \frac{\mathrm{d}\langle v_{\boldsymbol{k}_1}^s p_{\boldsymbol{k}_2}^{s'} \rangle}{\mathrm{d}\eta} &=  -i \langle [v_{\boldsymbol{k}_1}^sp_{\boldsymbol{k}_2}^{s'}, H_v] \rangle = \frac{-i}{2}\sum_{\lambda} \int \mathrm{d}^3 \boldsymbol{q} \langle v_{\boldsymbol{k}_1}^s[p_{\boldsymbol{k}_2}^{s'},H_v] + [v_{\boldsymbol{k}_1}^s,H_v]p_{\boldsymbol{k}_2}^{s'} \rangle \\
    &= -i \langle -i \omega^2 (k_2) v_{\boldsymbol{k}_1}^sv_{\boldsymbol{k}_2}^{s'} + ip_{\boldsymbol{k}_1}^s p_{\boldsymbol{k}_2}^{s'} \rangle = \langle p_{\boldsymbol{k}_1}^s p_{\boldsymbol{k}_2}^{s'}  \rangle -\omega^2 (k_2) \langle v_{\boldsymbol{k}_1}^s v_{\boldsymbol{k}_2}^{s'} \rangle,
\end{aligned}
\end{equation}
\begin{equation}\label{eq: comp 3rd observable}
\begin{aligned}
    \frac{\mathrm{d}\langle p_{\boldsymbol{k}_1}^s v_{\boldsymbol{k}_2}^{s'} \rangle}{\mathrm{d}\eta} &=  -i \langle [p_{\boldsymbol{k}_1}^sv_{\boldsymbol{k}_2}^{s'}, H_v] \rangle = \frac{-i}{2}\sum_{\lambda} \int \mathrm{d}^3 \boldsymbol{q} \langle p_{\boldsymbol{k}_1}^s[v_{\boldsymbol{k}_2}^{s'},H_v] + [p_{\boldsymbol{k}_1}^s,H_v]v_{\boldsymbol{k}_2}^{s'} \rangle \\
    &= -i \langle -i \omega^2 (k_1)v_{\boldsymbol{k}_1}^sv_{\boldsymbol{k}_2}^{s'} + ip_{\boldsymbol{k}_1}^s p_{\boldsymbol{k}_2}^{s'} \rangle = \langle p_{\boldsymbol{k}_1}^s p_{\boldsymbol{k}_2}^{s'}  \rangle -\omega^2 (k_1) \langle v_{\boldsymbol{k}_1}^s v_{\boldsymbol{k}_2}^{s'} \rangle .
\end{aligned}
\end{equation}
Lastly, we look at the $\langle p_{\boldsymbol{k}_1}^s p_{\boldsymbol{k}_2}^{s'} \rangle$ correlator. This is the most involved one, so we will compute the correlators separately.
\begin{equation}\label{eq:computation 4th oberservable}
\begin{aligned}
    &\frac{\mathrm{d}\langle p_{\boldsymbol{k}_1}^s p_{\boldsymbol{k}_2}^{s'} \rangle}{\mathrm{d}\eta} =  -i \langle [p_{\boldsymbol{k}_1}^sp_{\boldsymbol{k}_2}^{s'}, H_v] \rangle - \frac{\beta^2\gamma}{2(2\pi)^{3/2}} \sum_{\lambda,\lambda'} \cdot \\
    &\int \mathrm{d}^3\boldsymbol{k}\mathrm{d}^3\boldsymbol{n}_1\mathrm{d}^3\boldsymbol{n}_2 n_{1,l}(-k-n_1)^l n_{2,a} (k-n_2)^a \Tilde{C}_R (|\boldsymbol{k}|) \bigl \langle [p_{\boldsymbol{k}_1}^sp_{\boldsymbol{k}_2}^{s'},v_{\boldsymbol{n}_1}^{\lambda}v_{-\boldsymbol{k}-\boldsymbol{n}_1}^{\lambda}],v_{\boldsymbol{n}_2}^{\lambda'}v_{\boldsymbol{k}-\boldsymbol{n}_2}^{\lambda'}] \bigr \rangle.
\end{aligned}
\end{equation}
First we compute the commutator with the free Hamiltonian,
\begin{equation}\label{eq:computation 4th term, Hamiltonian term}
\begin{aligned}
    -i \langle [p_{\boldsymbol{k}_1}^sp_{\boldsymbol{k}_2}^{s'}, H_v] \rangle &= -i \langle p_{\boldsymbol{k}_1}^s[p_{\boldsymbol{k}_2}^{s'}, H_v] + [p_{\boldsymbol{k}_2}^{s'}, H_v]p_{\boldsymbol{k}_1}^s\rangle \\
    &= -i \langle -i \omega^2 (k_2)p_{\boldsymbol{k}_1}^sv_{\boldsymbol{k}_2}^{s'} -i \omega^2 (k_1) v_{\boldsymbol{k}_1}^s p_{\boldsymbol{k}_2}^{s'}\rangle \\
    &= - \omega^2 (k_2) \langle p_{\boldsymbol{k}_1}^sv_{\boldsymbol{k}_2}^{s'} \rangle - \omega^2 (k_1) \langle v_{\boldsymbol{k}_1}^s p_{\boldsymbol{k}_2}^{s'}\rangle.
\end{aligned}
\end{equation}
Second we work out the commutator with $v_{\boldsymbol{n}_1}^{\lambda}$
\begin{equation}
\begin{aligned}
    n_1^l& (-k-n_1)_l[p_{\boldsymbol{k}_1}^sp_{\boldsymbol{k}_2}^{s'},v_{\boldsymbol{n}_1}^{\lambda}v_{-\boldsymbol{k}-\boldsymbol{n}_1}^{\lambda}] \\
    &= n_1^l (-k-n_1)_l[p_{\boldsymbol{k}_1}^s[p_{\boldsymbol{k}_2}^{s'},v_{\boldsymbol{n}_1}^{\lambda}]v_{-\boldsymbol{k}-\boldsymbol{n}_1}^{\lambda} + p_{\boldsymbol{k}_1}^sv_{\boldsymbol{n}_1}^{\lambda} [p_{\boldsymbol{k}_2}^{s'},v_{-\boldsymbol{k}-\boldsymbol{n}_1}^{\lambda} ] \\
    &+ [p_{\boldsymbol{k}_1}^s,v_{\boldsymbol{n}_1}^{\lambda}]v_{-\boldsymbol{k}-\boldsymbol{n}_1}^{\lambda}p_{\boldsymbol{k}_2}^{s'} + v_{\boldsymbol{n}_1}^{\lambda}[p_{\boldsymbol{k}_1}^s,v_{-\boldsymbol{k}-\boldsymbol{n}_1}^{\lambda}]p_{\boldsymbol{k}_2}^{s'} ] \\
    &= n_1^l (-k-n_1)_l[p_{\boldsymbol{k}_1}^si\delta^{(3)}(\boldsymbol{k}_2+\boldsymbol{n}_1)\delta^{s'\lambda}v_{-\boldsymbol{k}-\boldsymbol{n}_1}^{\lambda} + p_{\boldsymbol{k}_1}^sv_{\boldsymbol{n}_1}^{\lambda} i\delta^{(3)}(\boldsymbol{k}_2-\boldsymbol{k}-\boldsymbol{n}_1)\delta^{s'\lambda} \\
    &+ i\delta^{(3)}(\boldsymbol{k}_1+\boldsymbol{n}_1)\delta^{s\lambda}v_{-\boldsymbol{k}-\boldsymbol{n}_1}^{\lambda}p_{\boldsymbol{k}_2}^{s'} + v_{\boldsymbol{n}_1}^{\lambda}i\delta^{(3)}(\boldsymbol{k}_1-\boldsymbol{k}-\boldsymbol{n}_1)\delta^{s\lambda}p_{\boldsymbol{k}_2}^{s'}] \\
    &= -i (-k_2)^l (-k+k_2)_l p_{\boldsymbol{k}_1}^s v_{-\boldsymbol{k}+\boldsymbol{k}_2}^{s'} -i (k_2-k)^l (-k_2)_l p_{\boldsymbol{k}_1}^s v_{-\boldsymbol{k}+\boldsymbol{k}_2}^{s'} \\
    &+ i (-k_1)^l (-k+k_1)_l  v_{-\boldsymbol{k}+\boldsymbol{k}_1}^sp_{\boldsymbol{k}_2}^{s'} + i  (-k+k_1)_l (-k_1)^l  v_{-\boldsymbol{k}+\boldsymbol{k}_1}^sp_{\boldsymbol{k}_2}^{s'} \\
    &= -2i \bigg( (-k_2)^l(-k+k_2)_l p_{\boldsymbol{k}_1}^s v_{-\boldsymbol{k}+\boldsymbol{k}_2}^{s'}  + v_{-\boldsymbol{k}+(-k_1)^l (-k+k_1)_l\boldsymbol{k}_1}^s p_{\boldsymbol{k}_2}^{s'}  \bigg).
\end{aligned}
\end{equation}
Now we plug this into the second commutator,
\begin{equation}
\begin{aligned}
    -i &n_2^a (k-n_2)_a \Big[ 2(-k_2)^l(-k+k_2)_l \Bigl \langle [p_{\boldsymbol{k}_1}^s, v_{\boldsymbol{n}_2}^{\lambda'}] v_{\boldsymbol{k}+\boldsymbol{n}_2}^{\lambda'} v_{-\boldsymbol{k}+\boldsymbol{k}_2}^{s'} + v_{\boldsymbol{n}_2}^{\lambda'} [p_{\boldsymbol{k}_1}^s, v_{\boldsymbol{k}-\boldsymbol{n}_2}^{\lambda'}] v_{-\boldsymbol{k}+\boldsymbol{k}_2}^{s'} \Bigr \rangle \\
    &+ 2(-k_1)^l (-k+k_1)_l \Bigl \langle v_{-\boldsymbol{k}+\boldsymbol{k}_1}^s [p_{\boldsymbol{k}_2}^{s'}, v_{\boldsymbol{n}_2}^{\lambda'}] v_{\boldsymbol{k}-\boldsymbol{n}_2}^{\lambda'} +  v_{-\boldsymbol{k}+\boldsymbol{k}_1}^s v_{\boldsymbol{n}_2}^{\lambda'} [p_{\boldsymbol{k}_2}^{s'}, v_{\boldsymbol{k}-\boldsymbol{n}_2}^{\lambda'}] \Bigr \rangle \Big] \\
    &= -i n_2^a (k-n_2)_a \Big[ 2(-k_2)^l(-k+k_2)_l \Bigl \langle -i \delta^{(3)}(\boldsymbol{k}_1+\boldsymbol{n}_2) \delta^{\lambda's} v_{\boldsymbol{k}+\boldsymbol{n}_2}^{\lambda'} v_{-\boldsymbol{k}+\boldsymbol{k}_2}^{s'} \\
    &- v_{\boldsymbol{n}_2}^{\lambda'} i \delta^{(3)}(\boldsymbol{k}_1+\boldsymbol{k}-\boldsymbol{n}_2) \delta^{\lambda's} v_{-\boldsymbol{k}+\boldsymbol{k}_2}^{s'} \Bigr \rangle \\
    &+ 2(-k_1)^l (-k+k_1)_l \Bigl \langle -v_{-\boldsymbol{k}+\boldsymbol{k}_1}^s i \delta^{(3)}(\boldsymbol{k}_2+\boldsymbol{n}_2) \delta^{\lambda's'} v_{\boldsymbol{k}-\boldsymbol{n}_2}^{\lambda'} -  v_{-\boldsymbol{k}+\boldsymbol{k}_1}^s v_{\boldsymbol{n}_2}^{\lambda'} i \delta^{(3)}(\boldsymbol{k}_2+\boldsymbol{k}-\boldsymbol{n}_2) \delta^{\lambda's'} \Bigr \rangle \Big] \\
    &= - \Bigl \langle 2(-k_2)^l(-k+k_2)_l 2(-k_1)^a(k+k_1)_a v_{\boldsymbol{k}+\boldsymbol{k}_1}^s v_{-\boldsymbol{k}+\boldsymbol{k}_2}^{s'}  \\
    &+ 2(-k_1)^l(-k+k_1)_l  2(-k_2)^a(k+k_2)_a v_{-\boldsymbol{k}+\boldsymbol{k}_1}^s v_{\boldsymbol{k}+\boldsymbol{k}_2}^{s'} \Bigr \rangle \\
    &= - 4 (-k_2)^l(-k+k_2)_l (-k_1)^a(k+k_1)_a \langle v_{\boldsymbol{k}+\boldsymbol{k}_1}^s v_{-\boldsymbol{k}+\boldsymbol{k}_2}^{s'} \rangle \\
    &- 4 (-k_1)^l(-k+k_1)_l (-k_2)^a(k+k_2)_a \langle v_{-\boldsymbol{k}+\boldsymbol{k}_1}^s v_{\boldsymbol{k}+\boldsymbol{k}_2}^{s'} \rangle,
\end{aligned}
\end{equation}
where we have used that the upper and lower indices are interchangeable because we sum over them. Now we use that we integrate over $\mathrm{d}^3\boldsymbol{k}$, so we can replace $\boldsymbol{k}$ with $-\boldsymbol{k}$ and add the two terms together. Plugging all this together into the final term of Eq. (\ref{eq:computation 4th oberservable}), and plugging (\ref{eq:computation 4th term, Hamiltonian term}) into the first term we get the final result
\begin{equation}\label{eq:comp 4th obersevable}
\begin{aligned}
    \frac{\mathrm{d}\langle p_{\boldsymbol{k}_1}^s p_{\boldsymbol{k}_2}^{s'} \rangle}{\mathrm{d}\eta} &= -\omega^2(k_2) \langle p_{\boldsymbol{k}_1}^s v_{\boldsymbol{k}_2}^{s'} \rangle -\omega^2(k_1) \langle v_{\boldsymbol{k}_1}^s p_{\boldsymbol{k}_2}^{s'}  \rangle \\
    &+ \frac{8\gamma\beta^2}{(2\pi)^{3/2}}\int \mathrm{d}^3\boldsymbol{k} k_1^l (k+k_1)_l k_2^a (k_2-k)_a \Tilde{C}_R(|\boldsymbol{k}|) \langle v_{\boldsymbol{k}+\boldsymbol{k}_1}^s v_{-\boldsymbol{k}+\boldsymbol{k}_2}^{s'}\rangle.
\end{aligned}
\end{equation}

These equations allow in particular to obtain predictions for the (cross-)correlations  
$\langle v_{\boldsymbol{k}_1}^s v_{\boldsymbol{k}_2}^{s'} \rangle$, between the two polarization states $+$ and $\times$, and hence to investigate the evolution of net polarization of the gravitational waves in terms of Stokes parameters $Q$, $U$ and $V$ (see, e.g.,~\cite{Smith:2016jqs,Thorne:2017jft} for the expression of the Stokes parameters in terms of the correlators of the two polarization states). The latter can be performed given some initial conditions and possibly given some properties of the environment. For example, we can see that a nonvanishing cross-correlation between the two polarization states $\times$ and $+$ will remain zero if the initial cross-correlation is zero. Similarly, one could think of initial conditions such that gravitational waves can be produced with only $+$ or with only $\times$ polarization states. Such initial conditions should be provided by a mechanism production, e.g. similar to the one envisaged in ~\cite{Tejerina-Perez:2024opu} (exploiting the interaction $\zeta\partial_{l}h_{ij}\partial_{l}h_{ij}$, which is similar to the kind of interactions we consider in this paper (see~\ref{GRcubic})).

\bibliographystyle{JHEP}
\bibliography{biblio.bib}

\end{document}